\documentclass[sn-mathphys-num]{sn-jnl}

\usepackage[utf8]{inputenc}
\usepackage{braket}

\usepackage{amsfonts, amssymb, amsmath, amsthm}
\usepackage{dsfont}
\usepackage{mathrsfs}
\usepackage{twemojis}
\usepackage{soul}

\usepackage{mathtools}
\usepackage[american]{babel}
\usepackage{xspace}

\usepackage{scalerel}

\usepackage{tikz}
\usetikzlibrary{shapes.geometric, positioning, cd}
\tikzcdset{
  cells={nodes={font=\small}}
}

\usepackage{tabularx}
\newcolumntype{C}{>{\centering\arraybackslash}X}
\usepackage{longtable}

\usepackage{xurl}
\hypersetup{breaklinks=true}

\usepackage{xr-hyper}

\usepackage{todonotes}

\usepackage{enumitem}

\usepackage{graphics}

\usepackage[linguistics]{forest}

\DeclareMathOperator{\Tr}{Tr}

\DeclareMathOperator{\Pos}{Pos}

\DeclareMathOperator{\Herm}{Herm}

\newcommand*{\Choi}{\ensuremath{\mathfrak{C}}\xspace}

\theoremstyle{plain}
\newtheorem{theorem}{Theorem}[section]
\newtheorem{lemma}[theorem]{Lemma}
\newtheorem{proposition}[theorem]{Proposition}

\theoremstyle{definition}
\newtheorem{definition}[theorem]{Definition}
\newtheorem{example}[theorem]{Example}

\theoremstyle{remark}
\newtheorem{remark}[theorem]{Remark}
\newtheorem{notation}[theorem]{Notation}

\newcommand*{\MyH}{\ensuremath{\mathcal{H}}\xspace}
\newcommand*{\Sys}[1]{\ensuremath{\mathrm{#1}}\xspace}
\newcommand*{\HilSpace}[1]{\ensuremath{\MyH_{#1}}\xspace}

\newcommand*{\LinearOps}{\ensuremath{\mathcal{L}}\xspace}
\newcommand*{\HermitianOps}[1]{\ensuremath{\Herm\left(#1\right)}\xspace}
\newcommand*{\PositiveOps}[1]{\ensuremath{\Pos\left(#1\right)}\xspace}

\newcommand*{\mydagger}{\scalerel*{\dag}{X}}
\newcommand*{\Adj}[1]{#1^{\mydagger}}

\newcommand*{\Trace}[1]{\ensuremath{\Tr\!\left[#1\right]}\xspace}

\DeclareMathOperator{\Dimension}{dim}

\newcommand*{\DefnFont}[1]{\emph{#1}}

\newcommand*{\alphabet}{\ensuremath{\mathbb{A}}\xspace}
\newcommand*{\ty}{\ensuremath{{\mathtt{Types}}_{\alphabet}}\xspace}
\newcommand*{\et}{\ensuremath{{\mathtt{EleTypes}}_{\alphabet}}\xspace}

\newcommand*{\tystruct}[1]{\ensuremath{\textup{\texttt{st}}(#1)}\xspace}
\newcommand*{\tyord}[1]{\ensuremath{\textup{\texttt{ord}}(#1)}\xspace}

\newcommand*{\tyto}{\ensuremath{\to}\xspace}
\newcommand*{\partype}{\ensuremath{\mathbin{\scaleobj{0.8}{\boxtimes}}}\xspace}
\newcommand*{\parmap}{\ensuremath{\mathbin{\scaleobj{0.8}{\boxtimes}}}\xspace}
\newcommand*{\id}[1]{\ensuremath{\mathcal{I}_{#1}}\xspace}

\newcommand*{\tyin}[1]{\ensuremath{\texttt{in}(#1)}\xspace}
\newcommand*{\tyout}[1]{\ensuremath{\texttt{out}(#1)}\xspace}

\newcommand*{\LinearMaps}{\ensuremath{\textsf{L}}\xspace}
\newcommand*{\HMaps}{\ensuremath{\textsf{H}}\xspace}
\newcommand*{\KMaps}{\ensuremath{\textsf{K}}\xspace}
\newcommand*{\LeftIP}{\langle}
\newcommand*{\RightIP}{\rangle}

\newcommand*{\InductionIndex}{\ensuremath{n}\xspace}

\newcommand*{\arbcone}{\Gamma}

\newcommand*{\higherfont}{\mathcal}

\newcommand*{\mylabel}[2]{#2\def\@currentlabel{#2}\label{#1}}

\usepackage{csquotes}

\renewcommand*{\tyto}{\ensuremath{\to}\xspace}
\renewcommand*{\tyin}[1]{\ensuremath{\textrm{in}(#1)}\xspace}
\renewcommand*{\tyout}[1]{\ensuremath{\textrm{out}(#1)}\xspace}
\renewcommand*{\tystruct}[1]{\ensuremath{\textup{\textrm{st}}(#1)}\xspace}
\renewcommand*{\tyord}[1]{\ensuremath{\textup{\textrm{ord}}(#1)}\xspace}

\begin{document}

 \title{Towards the simulation of higher-order quantum resources: a general type-theoretic approach}
 \author[1,2,3]{\fnm{Samuel B.} \sur{Steakley}}\email{samuel.steakley@taltech.ee}
 \author[1,2]{\fnm{Elia} \sur{Zanoni}}\email{elia.zanoni94@gmail.com}
 \author[1,2]{\fnm{Carlo Maria} \sur{Scandolo}}\email{carlomaria.scandolo@ucalgary.ca}
 
 \affil[1]{\orgdiv{Department of Mathematics \& Statistics}, \orgname{University of Calgary}, \orgaddress{\city{Calgary}, \postcode{T2N 1N4}, \state{AB}, \country{Canada}}}
 \affil[2]{\orgdiv{Institute for Quantum Science and Technology}, \orgname{University of Calgary}, \orgaddress{\city{Calgary}, \postcode{T2N 1N4}, \state{AB}, \country{Canada}}}
  \affil[3]{\orgdiv{Department of Software Science}, \orgname{Tallinn University of Technology}, \orgaddress{\city{Tallinn}, \postcode{19086}, \country{Estonia}}}


\abstract{Quantum resources exist in a hierarchy of multiple levels.
  At order zero, quantum states are transformed by linear maps (channels, or gates) in order to perform computations or simulate other states.
  At order one, gates and channels are transformed by linear maps (superchannels) in order to simulate other gates.
  To develop a full hierarchy of quantum resources, beyond those first two orders, and to account for the fact that quantum protocols can interconvert resources of different orders, we need a theoretical framework that addresses all orders in a uniform manner.
  We introduce a framework based on a system of types, which label the different kinds of objects that are present at different orders.
  We equip the framework with a parallel product operation that modifies and generalizes the tensor product so as to be operationally meaningful for maps of distinct and arbitrary orders.
  Finally, we introduce a family of convex cones that generalize the notion of complete positivity to all orders, with the aim of characterizing the objects that are physically admissible,
  facilitating an operational treatment of quantum objects at any order.}

\maketitle

\section{Introduction}



Quantum science and technology have been the subject of such interest in recent years \cite{knightUKNationalQuantum2019,raymerUSNationalQuantum2019,sussmanQuantumCanada2019,kochQuantumOptimalControl2022} that this period has been dubbed the second quantum revolution \cite{dowlingQuantumTechnologySecond2003}.
The power of quantum information processing is being investigated in various areas, but especially quantum sensing \cite{degenQuantumSensing2017}, quantum computing \cite{aaronsonFutureQuantumComputing2025,sandersSuccessFailureQuantum2025}, and quantum communications \cite{sidhuAdvancesSpaceQuantum2021}.
The overarching theme is that quantum technologies and devices constitute a resource that confers a unique quantum advantage in information processing.
Therefore, it is imperative to develop a scientific theory that precisely captures the notion of quantum resource.
Indeed, we must seek to understand the general laws of how quantum resources can be generated, interconverted, and optimally exploited \cite{chitambarQuantumResourceTheories2019,gourQuantumResourceTheories2025}.


In the hierarchy of different kinds of quantum resources, the base order (i.e.\ level)
consists of quantum states, with valuable properties such as superposition \cite{theurerResourceTheorySuperposition2017} and coherence \cite{streltsovColloquiumQuantumCoherence2017}, being manipulated by transformations known as quantum channels.
These states are called \DefnFont{static resources} \cite{chitambarQuantumResourceTheories2019,gourQuantumResourceTheories2025} and they are e.g.\ the subject of quantum computation:
we use quantum gates to manipulate input states in order to create superpositions and entanglement, and to run quantum algorithms.
We refer to this order of the hierarchy as order zero.


At the next order of the hierarchy, we consider e.g.\ quantum gates.
In general, the quantum operations that perform the basic steps of an algorithm are not all equally easy to implement \cite{gottesmanHeisenbergRepresentationQuantum1998,veitchResourceTheoryStabilizer2014},
and so for efficiency, it may become necessary to perform optimizations on a given quantum circuit.
For instance, we may wish to reduce a circuit's \( T \)-count \cite{heyfronEfficientQuantumCompiler2018}, i.e.\ the number of \( T \) gates it contains.
The general issue here is the task of circuit transpilation or optimization: an algorithm can be accomplished using a given collection of gates, but we wish to simulate those gates using another, less costly collection.
We encounter a similar problem when we face a quantum communication task in which the given quantum channels may be noisy, in which case we want a protocol that utilizes the given quantum channels in order to simulate a quantum channel with as little noise as possible.
A solution to either task is a protocol that transforms an input channel into an output channel, and which is necessarily a kind of quantum transformation known as a \DefnFont{superchannel} \cite{chiribellaTransformingQuantumOperations2008,gourComparisonQuantumChannels2019}.
This forms order one of the hierarchy of quantum resources: quantum channels as objects being manipulated by superchannels.
In this context, valuable channels are known as \DefnFont{dynamical resources} \cite{chitambarQuantumResourceTheories2019,gourDynamicalResources2020,gourHowQuantifyDynamical2019,liuResourceTheoriesQuantum2019,liuOperationalResourceTheory2020,yuanOneshotDynamicalResource2020}.


Although static and dynamical resources are of distinct kinds, 
they are not separate in a practical sense, because 
certain quantum protocols
make it possible to use a static resource to simulate a dynamical resource, or \textit{vice versa}.
In effect, such a protocol allows one kind of resource to be converted into the other.
For example, this is the case in the teleportation protocol \cite{bennettTeleportingUnknownQuantum1993}, in which one entangled state is converted into a noiseless channel.
But there are many other such protocols \cite{devetakDistillingCommonRandomness2004,devetakResourceFrameworkQuantum2008}, such as entanglement distribution, in which a quantum channel is turned into a quantum entangled state.
This leads to inevitable tradeoffs between static and dynamical resources in information processing tasks, and so we need theoretical methods that treat them on a common ground, instead of separate methodologies.


Static and dynamical resources are only two orders of a much larger hierarchy.
The same reasoning that leads us to consider the manipulation of channels by superchannels also invites us to consider all feasible ways of manipulating superchannels.
This leads to order two of the hierarchy: superchannels as objects being transformed by an even ``higher'' form of quantum protocol (for lack of a commonly accepted name, and despite the awkwardness, these could be called ``supersuperchannels'').
In fact, there is no theoretical reason to stop at any particular order:
for any number \( n \), we may define order \( n + 1 \) by making its objects the transformations of order \( n \), and making its transformations all feasible ways of manipulating those objects.
Thus, the full hierarchy includes infinitely many orders.
In order to capture the full possibilities of quantum protocols, we need a theoretical methodology that applies to protocols at all orders.

By investigating the orders above one, researchers have uncovered quantum protocols that exhibit unexpected and valuable features, such as the quantum SWITCH \cite{ chiribellaQuantumComputationsDefinite2013, colnaghiQuantumComputationProgrammable2012, chiribellaPerfectDiscriminationNosignalling2012, chiribellaQuantumShannonTheory2019, kristjanssonResourceTheoriesCommunication2020, kristjanssonSecondquantisedShannonTheory2022} with its feature of indefinite causal order \cite{oreshkovQuantumCorrelationsNo2012, oreshkovCausalCausallySeparable2016,baumelerSpaceLogicallyConsistent2016,wechsQuantumCircuitsClassical2021}.
It has been found that the SWITCH
speeds up certain information processing tasks \cite{colnaghiQuantumComputationProgrammable2012, chiribellaPerfectDiscriminationNosignalling2012,chiribellaQuantumShannonTheory2019,kristjanssonResourceTheoriesCommunication2020,kristjanssonSecondquantisedShannonTheory2022,zhaoQuantumMetrologyIndefinite2020},
and thus we should view these ``higher'' protocols as yet another source of quantum advantage.
There are reasons for interest for fundamental science too, since
the phenomenon of indefinite causal order has been theoretically linked to
quantum gravity.
The idea is
that if quantum superpositions of spacetime occur at the quantum gravity scale,
then the causal structure of local light cones may itself be in superposition, hence indefinite.
Studying higher quantum protocols could provide us with candidate models for indefinite causal order at the level of spacetime.

We propose a theoretical framework that addresses both of the aforementioned desiderata:
all possible orders of the hierarchy are included, and they are all unified in the sense that we can consider all ways of combining, simulating, and interconverting objects from arbitrary orders.
This is all to say that our formalism addresses the full scope of what is known as higher-order quantum theory.

Our framework is partly modeled on the one discussed in \cite{perinottiCausalStructuresClassification2017,bisioTheoreticalFrameworkHigherorder2019, apadulaNosignallingConstrainsQuantum2024} (see \cite{kissingerCategoricalSemanticsCausal2019,simmonsHigherOrderCausalTheories2022b,Simmonslogic,HeffordWilson,WilsonPhD,wilson2022freepolycategoriesunitarysupermaps,WilsonCausal,Supermapslocality} for related approaches).
Like theirs, it addresses higher-order quantum theory for finite-dimensional systems, and it identifies and differentiates the different kinds of higher-order transformation with the aid of a simple type system.
However, we introduce a novel algebraic method for describing the parallel application of two higher-order maps of any arbitrary orders.
As a further technical point, we opt against using the Choi representation to define higher-order quantum objects.
One advantage of this is that we avoid the ambiguities of which systems to associate with different entries of a concrete Choi matrix,
as highlighted in \cite{zanoniChoidefinedResourceTheories2025}.

\subsection{Content overview}
We first build a set of types, \ty, out of a given ``base'' set \alphabet of letters indicating physical systems.
Types may be thought of as labels that we use to differentiate and identify all the various kinds of objects (states and higher-order transformations) that arise in higher-order quantum theory.
Using a simple inductive construction, we associate each type \( x \in \ty \) with a Hilbert space \( \LinearMaps(x) \) of linear maps, which we call the space of linear maps of type \( x \).
The Hilbert spaces \( \LinearMaps(x) \) provide a universe within which the physically meaningful maps (states, channels, superchannels, etc.) are to be characterized.

We then introduce an algebraic operation on linear maps, which we call the \DefnFont{parallel product} and denote by \parmap.
Given a pair of types \( {x,y\in \ty} \) as well as a map \( \higherfont{M} \) of type \( x \) and a map \( \higherfont{N} \) of type \( y \),
we define a parallel product type \( x \partype y \in \ty \) and a parallel product map \( \higherfont{M} \parmap \higherfont{N} \) of type \( x \partype y \).
The parallel product is a generalization of the notion of ``extended event'' defined in \cite[Definition~4.3]{bisioTheoreticalFrameworkHigherorder2019}
, in that the parallel product allows maps of all types to be combined, rather than only allowing the extension with a base type.
The motivating example for the parallel product is the tensor product of channels.
We will see that the tensor product of two higher-order transformations of different kinds is \emph{not} suitable as a parallel product of the two, but that a general parallel product operation can be built by using the tensor product within a simple inductive formula.


Finally, we take a first step toward the general problem of characterizing the subset of physically meaningful maps within each set \( \LinearMaps(x) \).
Our point of departure consists of the fact that only the positive semidefinite operators are used to describe states of a quantum system,
as well as the consequent facts that channels must be completely positive (or, ``CP'') and that superchannels must be completely CP-preserving \cite{gourComparisonQuantumChannels2019,burnistonNecessarySufficientConditions2020}.
Extending the same logic, a physically meaningful map of any type \( x \) should satisfy a certain generalized property of complete positivity, and this is what we define:
for each \( x \in \ty \) we define a subset \( \KMaps(x) \subset \LinearMaps(x) \), hence a property of linear maps of type \( x \), that inductively generalizes the notions of completely positive map and completely CP-preserving map to all types.
We show that each \( \KMaps(x) \) forms a convex cone, just like the subset of positive semidefinite operators does.

The rest of the paper is organized as follows.
In Section~\ref{sec:typesandlinearmaps}, we introduce our type system and we provide a convenient graphical representation in terms of trees.
Then we define the \LinearMaps family of spaces of higher-order linear maps.
In Section~\ref{sec:parallelproduct}, we introduce the parallel product of types and maps, again with a convenient interpretation in terms of trees.
In Section~\ref{sec:completelypositivemaps}, we introduce a higher-order generalization of complete positivity.
Conclusions are drawn in Section~\ref{sec:conclusion}.

\section{Types and linear maps} \label{sec:typesandlinearmaps}

The first main task in this section is to construct a set \ty of types.
For our purposes, types may be thought of as \emph{labels} that we use to identify and differentiate all the different kinds of objects that arise in higher-order quantum theory.

At the most basic level, we assume that we are interested in studying some particular quantum system and that we have chosen a label to refer to that system.
For example, if we wish to consider a generic 2-dimensional quantum system, then we may label this system by the letter \Sys{B}, and we may choose the Hilbert space \( \HilSpace{\Sys{B}} = \mathbb{C}^2 \) to model it.
In general, we let \alphabet be a set of labels referring to the quantum systems we wish to study,
and these labels form the base of our type system:
for each \( A \in \alphabet \), states of system \Sys{A} are \DefnFont{states of type}~\Sys{A}.

For any pair of systems, there is a corresponding composite system. We assign the composite system a type by combining the two labels into a ``word'':
for each \( \Sys{A}, \Sys{B} \in \alphabet \), states of the corresponding composite system are \DefnFont{states of type} \Sys{AB}.
Composite systems with three or more components are handled in the same way; e.g.\ for any three systems \( \Sys{A}, \Sys{B}, \Sys{C} \in \alphabet \), states of the corresponding composite system are \DefnFont{states of type} \Sys{ABC}.

In order to describe transformations, rather than states, we introduce a new symbol into our types: the arrow symbol \tyto.
For example, the type of a channel consists of an arrow with an input system on its left and an output system on its right:
for each \( \Sys{A}, \Sys{B} \in \alphabet \), channels with input \Sys{A} and output \Sys{B} are \DefnFont{channels of type} \( \Sys{A} \tyto \Sys{B} \).
By extension, a generic superchannel \cite{chiribellaTransformingQuantumOperations2008,gourComparisonQuantumChannels2019} would be of type \( {(\Sys{A} \tyto \Sys{B}) \tyto (\Sys{C} \tyto \Sys{D})} \), and a generic transformation that takes a channel as input and returns a superchannel as output would be of type~\( {(\Sys{A} \tyto \Sys{B}) \tyto ((\Sys{C} \tyto \Sys{D}) \tyto (\Sys{E} \tyto \Sys{F}))} \).
(Notice the use of parentheses to separate arrow symbols, which is necessary in order to disambiguate the ``order of inputs and outputs.'').
The full set contains infinitely many types, for describing the infinitely many kinds of higher-order transformations.

The second main task is to put the set \ty of types to its first major use:
to inductively define a finite-dimensional Hilbert space \( \LinearMaps(x) \), for each \( x \in \ty \), whose elements we call the \DefnFont{linear maps of type} \( x \).
Informally, we refer to \LinearMaps as a \emph{family} of Hilbert spaces, because it consists of one space for each type.
The \LinearMaps family provides us with a universe of higher-order linear maps within which the physically meaningful maps are to be characterized.

\subsection{Types}

Let \alphabet be a set, whose elements we think of as labels for physical systems.
We now present a construction that generalizes the former motivating examples into a full set of types generated by \alphabet.

First we construct the set of elementary types.
These types describe all base systems, all composites of base systems, and also the ``trivial'' system (to be explained momentarily).
\begin{definition} \label{def:elementarytypes}
  \et is the set of all finite-length words made of elements of \alphabet.
\end{definition}
``Word'' simply means sequence of symbols.
For each base system \( \Sys{A} \in \alphabet \), we identify the label \Sys{A} with the length-one sequence \( \Sys{A} \in \et \).
Composite systems are described by sequences of length greater than one,
including repeated instances of the same system;
e.g.\ \( \Sys{A},\Sys{B},\Sys{C}\in \alphabet \) entails \( \Sys{AABCB} \in \et \).

The elementary types also include a unique sequence of length zero (hence, \et is non-empty even if \alphabet is empty).
We call this type the \DefnFont{trivial type}, and we denote it by \( \Sys{I} \) (in particular,
we assume that \Sys{I} is \emph{not} an element of \alphabet).
We use the trivial type to refer to the \DefnFont{trivial system}, which is a sort of fictitious system that indicates the lack of a physical system, or ``nothing.''

The full set of types is constructed by inductively expanding \et with all ``arrow types.''
These additional types, formally called ``non-elementary types,'' are how we describe objects of higher order than states, e.g.\ channels, superchannels, etc.
\begin{definition} \label{def:types}
  \ty is the smallest set of formal expressions satisfying the following two rules:
  \begin{gather}
    \label{eq:typesrule1} \et \subseteq \ty \tag{R1} \\[0.5em] 
    \label{eq:typesrule2} \text{If } x,y\in\ty \text{, then } x \tyto y \in \ty \tag{R2}
  \end{gather}
\end{definition}
The notation \( x \tyto y \) denotes the concatenation of the expressions \( x \) and \( y \), with an arrow symbol placed in between;
also, round brackets must be placed around \( x \) (respectively, around \( y \)) if \( x \) (resp.\ \( y \)) contains an arrow symbol.
The purpose of the brackets is to distinguish input from outputs.
E.g.\ if \( x = \Sys{AB} \) and \( y = \Sys{C} \) then \( x \tyto y = \Sys{AB} \tyto \Sys{C} \), but if \( y = \Sys{B} \tyto \Sys{C} \) then \( {x \tyto y = \Sys{AB} \tyto (\Sys{C} \tyto \Sys{D})} \). Elementary types and arrows alone cannot give a full account of a type: \( {\Sys{AB} \tyto (\Sys{C} \tyto \Sys{D})} \) and \( {(\Sys{AB} \tyto \Sys{C}) \tyto \Sys{D}} \) are two very different types, despite having the same elementary types and the arrows positioned in the same places. The former corresponds to the physical situation in which a quantum bipartite state of $ {\Sys{AB}}$ is transformed into a quantum channel from $ {\Sys{C}}$ to $ {\Sys{D}}$. The latter corresponds to the physical situation in which a quantum channel from $ {\Sys{AB}}$ to  $ {\Sys{C}}$ is transformed into a quantum state of $ {\Sys{D}}$.

Finally we note that, due to rule~\eqref{eq:typesrule2}, \ty is necessarily an infinite set.
Even if \alphabet is empty, \et contains the trivial type (i.e.\ the empty sequence), denoted \( \Sys{I} \).
Then \( {\Sys{I} \in \ty} \) by~\eqref{eq:typesrule1}.
Then, by repeated application of~\eqref{eq:typesrule2}, it follows that
\( \Sys{I} \tyto \Sys{I} \in \ty \),
and \( \Sys{I} \tyto (\Sys{I} \tyto \Sys{I}) \in \ty \),
and \( {\Sys{I} \tyto (\Sys{I} \tyto (\Sys{I} \tyto \Sys{I})) \in \ty} \),
and so on.

\subsubsection{Types as trees}

In the fields of mathematics that study formal languages (e.g.\ mathematical logic), the set \ty would be recognized as a simple kind of formal language.
We have chosen to introduce types as ``formal expressions,'' by which we mean sequences of symbols, but it is well-known that the elements of this kind of language can be formalized in more than one useful way.
They can also be formalized as \emph{syntax trees}.
It would take us too far afield to go over the precise details, but we can still profit from visualizing a type in two equivalent ways: as a formal expression or as a syntax tree.
This is best demonstrated by way of example.

A syntax tree is a certain kind of tree, which means that it consists of \DefnFont{vertices} that are connected by \DefnFont{edges} (an edge is drawn as a line from one vertex to another).
Suppose \( \Sys{A},\Sys{B},\Sys{C},... \) are elements of \alphabet.
When we render the type \( \Sys{A} \tyto \Sys{B} \) as a syntax tree, as in~\eqref{ex:treeexample1}, this is done as follows.
Each of the three symbols becomes a vertex.
We place the \tyto vertex at the top, and we call this vertex the \DefnFont{root} of the tree.
Finally, we add edges connecting \Sys{A} and \Sys{B} to the root, placing \Sys{A} below left and \Sys{B} below right.
\begin{equation} \label{ex:treeexample1}
  \Sys{A} \tyto \Sys{B}
  \quad \leftrightsquigarrow \quad
  \begin{tikzcd}[row sep={1em}, column sep={1.6em, between origins}]
    & \tyto & \\
    \Sys{A} & & \Sys{B}
    \arrow[from=1-2, to=2-1, dash] \arrow[from=1-2, to=2-3, dash]
  \end{tikzcd}
\end{equation}

For the type \( (\Sys{A} \tyto \Sys{B}) \tyto (\Sys{C} \tyto \Sys{D}) \), we apply a similar procedure, but now in multiple steps proceeding from the bottom up.
We begin with the \tyto symbols which are the ``deepest'' with respect to nesting of parentheses.
By this we mean each \tyto symbol meeting the following description:
a \tyto symbol such that the symbols to its immediate left and right are both letters (elements of \alphabet), as opposed to parenthesis symbols.
In the type \( (\Sys{A} \tyto \Sys{B}) \tyto (\Sys{C} \tyto \Sys{D}) \), there are two deepest \tyto symbols: the \tyto with \Sys{A} on its left and \Sys{B} on its right, and the \tyto with \Sys{C} on its left and \Sys{D} on its right.
We take these two and build their sub-trees according to the same procedure described above, which results in the following two trees:
\begin{align}
  \Sys{A} \tyto \Sys{B}
  \quad \leftrightsquigarrow \quad
  \begin{tikzcd}[row sep={1em}, column sep={1.6em, between origins}, ampersand replacement=\&]
    \& \tyto \& \\
    \Sys{A} \& \& \Sys{B}
    \arrow[from=1-2, to=2-1, dash] \arrow[from=1-2, to=2-3, dash]
  \end{tikzcd}
  &&
  \Sys{C} \tyto \Sys{D}
  \quad \leftrightsquigarrow \quad
  \begin{tikzcd}[row sep={1em}, column sep={1.6em, between origins}, ampersand replacement=\&]
    \& \tyto \& \\
    \Sys{C} \& \& \Sys{D}
    \arrow[from=1-2, to=2-1, dash] \arrow[from=1-2, to=2-3, dash]
  \end{tikzcd}
\end{align}
Finally, we form one big tree by placing the outermost \tyto at the top and joining the two sub-trees underneath it, with an edge from the overall root to the root of each sub-tree:
\begin{equation}
  (\Sys{A} \tyto \Sys{B}) \tyto (\Sys{C} \tyto \Sys{D})
  \quad \leftrightsquigarrow \quad
  \begin{tikzcd}[row sep={1em}, column sep={1.6em, between origins}]
    & & & \tyto & & & \\
    & \tyto & & & & \tyto & \\
    \Sys{A} & & \Sys{B} & & \Sys{C} & & \Sys{D}
    \arrow[from=1-4, to=2-2, dash] \arrow[from=1-4, to=2-6, dash] \arrow[from=2-2, to=3-1, dash] \arrow[from=2-2, to=3-3, dash] \arrow[from=2-6, to=3-5, dash] \arrow[from=2-6, to=3-7, dash]
  \end{tikzcd}
\end{equation}

Extending the same ``bottom-up'' procedure, one can translate any type to a tree.
We provide a few more examples:
\begin{equation}
  (\Sys{A} \tyto \Sys{BA}) \tyto \Sys{ABC}
  \quad \leftrightsquigarrow \quad
  \begin{tikzcd}[row sep={1em}, column sep={2em, between origins}, ampersand replacement=\&]
    \& \& \ar[dl, dash] \tyto \ar[dr, dash] \& \\
    \& \ar[dl, dash] \tyto \ar[dr, dash] \& \& \Sys{ABC} \\
    \Sys{A} \& \& \Sys{BA} \&
  \end{tikzcd}
\end{equation}
\begin{equation}
  \Sys{ABC} \tyto (\Sys{A} \tyto \Sys{BA})
  \quad \leftrightsquigarrow \quad
  \begin{tikzcd}[row sep={1em}, column sep={2em, between origins}, ampersand replacement=\&]
    \& \ar[dl, dash] \tyto \ar[dr, dash] \& \& \\
    \Sys{ABC} \& \& \ar[dl, dash] \tyto \ar[dr, dash] \& \\
    \& \Sys{A} \& \& \Sys{BA}
  \end{tikzcd}
\end{equation}
\begin{equation}
  (\Sys{AB} \tyto \Sys{C}) \tyto (\Sys{DEF} \tyto (\Sys{GH} \tyto \Sys{K}))
  \quad \leftrightsquigarrow \quad
  \begin{tikzcd}[row sep={1em}, column sep={0em}, ampersand replacement=\&]
    \& \& \& \arrow[dll, dash] \tyto \arrow[dr, dash] \& \& \& \& \\
    \& \arrow[dl, dash] \tyto \arrow[dr, dash] \& \& \& \arrow[dl, dash] \tyto \arrow[dr, dash] \& \& \\
    \Sys{AB} \& \& \Sys{C} \& \Sys{DEF} \& \& \arrow[dl, dash] \tyto \arrow[dr, dash] \& \\
    \& \& \& \& \Sys{GH} \& \& \Sys{K}
  \end{tikzcd}
\end{equation}

Through the correspondence that these examples illustrate, one sees that in a tree there is no need for parentheses because the edges between \tyto vertices make the ``order of operations'' clear.
And for large types, a tree is more legible because it is laid out in a 2-dimensional form, instead of a mess of brackets.

\subsubsection{Basic properties of types}

It is known that the set \ty, by being constructed in the manner of Definition~\ref{def:types},
has the property of being freely inductively generated
with respect to rules~\eqref{eq:typesrule1} and~\eqref{eq:typesrule2}.
For background on this fact and the precise meaning of ``freely inductively generated,'' we refer the reader to \cite{avigadMathematicalLogicComputation2022}.
However, we are mainly interested in two consequences of this.

First, there is Lemma~\ref{lem:uniquereadability}, which is a close corollary of the fact that \ty is freely inductively generated with respect to rules~\eqref{eq:typesrule1} and~\eqref{eq:typesrule2}, and which provides a simple way of understanding all of the elements of \ty by dividing them into two distinct categories.
A type \( x \) must either be elementary, or else be the arrow type \( a \tyto b \) for some unique choice of types \( a \) and \( b \) (which clearly must be ``smaller'' than \( x \), i.e.\ with fewer symbols).
\begin{lemma} \label{lem:uniquereadability}
  Every \( x \in \ty \) satisfies exactly one of the following statements:
  \begin{enumerate}[label=(\roman*)]
    \item \( x \in \et \).
    \item \( x = a \tyto b \) for a unique pair \( a,b\in\ty \).
  \end{enumerate}
\end{lemma}
We need not dwell on the details, but for the interested reader a proof is provided in the appendix.
One can easily judge which of the two above cases a given type falls into:
either the type contains no \tyto symbol, or else there is exactly one \tyto symbol on the outside of all brackets.
In that case, \( a \) is the type on the left of the outermost \tyto, and \( b \) is the type on the right.
When viewing a type as a tree, one instead finds the uppermost \tyto symbol, in which case \( a \) is the type given by the full sub-tree attached to the root's left-hand edge, and \( b \) is the one attached to the right-hand edge. 

As a consequence of Lemma~\ref{lem:uniquereadability}, each non-elementary type has a uniquely defined input type and output type.
\begin{definition}
  Given \( x = a \tyto b \in \ty \), the \DefnFont{input type of} \( x \) is defined
  \begin{equation}
    \tyin{x} = a ,
  \end{equation}
  and the \DefnFont{output type of} \( x \) is defined
  \begin{equation}
    \tyout{x} = b .
  \end{equation}
\end{definition}

The second important consequence that we take from \ty being freely inductively generated is that it is valid to carry out proofs and more general constructions on \ty by induction over rules~\eqref{eq:typesrule1} and~\eqref{eq:typesrule2}.
We make constant use of these techniques throughout this work.
Definition~\ref{def:typeorder} provides a first, simple example.

The term ``order,'' which we have used informally up to this point, is now defined as a property of types, describing the degree of ``nestedness'' of arrow symbols in a type, or equivalently describing the height of the syntax tree.
A simple inductive formula is used.
\begin{definition} \label{def:typeorder}
  Let \( x \in \ty \).
  The \DefnFont{order} of \( x \) is defined:
  \begin{equation} \label{eq:typeorderrecursiveformula}
    \tyord{x} =
    \begin{cases*}
      0 \enspace & if \( x \in \et \)  \\
      1 + \max\{\tyord{a}, \tyord{b}\} \enspace & if \( x = a \tyto b \) 
    \end{cases*}
  \end{equation}
\end{definition}

We demonstrate how to calculate the order of a type in the following example.
\begin{example} \label{ex:typeorder}
  Let \( x = (\Sys{AB} \tyto \Sys{C}) \tyto (\Sys{DEF} \tyto (\Sys{GH} \tyto \Sys{K})) \). One way to compute the order of \( x \) is to apply Definition~\ref{def:typeorder} in an algorithmic manner, compute the order ``from the bottom up.''
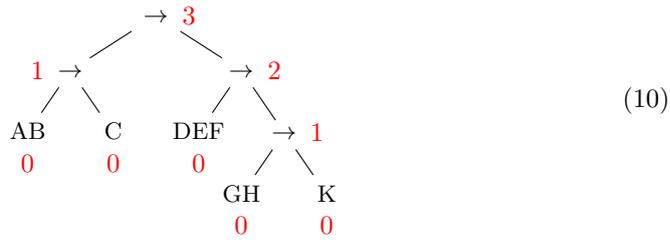
  \begin{equation}
    \begin{tikzcd}[row sep={1em}, column sep={1.6em, between origins}]
      & & & \arrow[dll, dash] \tyto \ar[phantom, "3"{right=.75em, red}] \arrow[drr, dash] & & & & \\
      & \arrow[dl, dash] \tyto \ar[phantom, "1"{left=.75em, red}] \arrow[dr, dash] & & & & \arrow[dl, dash] \tyto \ar[phantom, "2"{right=.75em, red}] \arrow[dr, dash] & & \\
      \ar[phantom, "0"{below=.6em, red}] \Sys{A}\Sys{B} & & \Sys{C} \ar[phantom, "0"{below=.6em, red}] & & \ar[phantom, "0"{below=.6em, red}] \Sys{D}\Sys{E}\Sys{F} & & \arrow[dl, dash] \tyto \ar[phantom, "1"{right=.75em, red}] \arrow[dr, dash] & \\
      & & & & & \ar[phantom, "0"{below=.6em, red}] \Sys{G}\Sys{H} & & \Sys{K} \ar[phantom, "0"{below=.6em, red}]
    \end{tikzcd}
  \end{equation}
  In this figure we have annotated in red the result of assigning zero to the elementary types in the tree, and assigning to each \tyto one plus the maximum of the numbers at the vertices immediately beneath it.
  This shows that \( \tyord{x} = 3 \)
  The same result is also given, in complete generality, by finding the maximum length of all paths (with non-repeat edges) from the root vertex to one of the bottom vertices.
  In the case of \( x \), this path is indicated in red in the figure below:
  
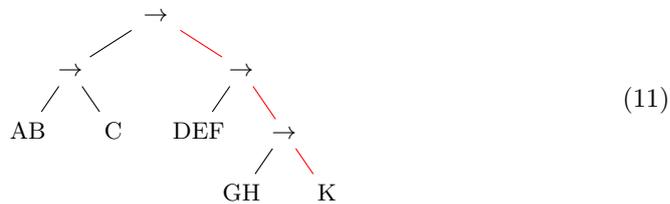
\begin{equation}
    \begin{tikzcd}[row sep={1em}, column sep={1.6em, between origins}]
      & & & \arrow[dll, dash] \tyto \arrow[drr, dash, red] & & & & \\
      & \arrow[dl, dash] \tyto \arrow[dr, dash] & & & & \arrow[dl, dash] \tyto \arrow[dr, dash, red] & & \\
      \Sys{A}\Sys{B} & & \Sys{C} & & \Sys{D}\Sys{E}\Sys{F} & & \arrow[dl, dash] \tyto \arrow[dr, dash, red] & \\
      & & & & & \Sys{G}\Sys{H} & & \Sys{K}
    \end{tikzcd}
  \end{equation}
  Because of this interpretation, this quantity is sometimes known as the height of the tree.
\end{example}

Elementary types can now be characterized as being precisely the types of order zero.
Types of order one are exactly the non-elementary types whose input and output types are both of order zero.
Types of order two are exactly the non-elementary types whose input and output types are both of order no greater than one, but not both of order zero.
And so on and so forth.

By the same token, states are described by types of order zero, channels are described by types of order one, and superchannels are described by types of order two.
However, types of order two do not exclusively describe superchannels.
A transformation that sends an input state to an output channel would be of type e.g.\ \( \Sys{A} \tyto (\Sys{B} \tyto \Sys{C}) \), which is of order two.
Conversely, a transformation that sends an input channel to an output state would also be described by a type of order two.

The structure of a type is a property that corresponds to the kind of object the type describes irrespective of the particular physical systems the type involves.
By ``kind of object'' we mean state, channel, or superchannel, etc.
We use the syntax tree perspective on types to give a clear, simple definition of structure.
Notice that in all our examples of types as trees, each vertex of the syntax tree is marked with some label, which is either the \tyto symbol or some elementary type.
Mathematically speaking, labels are not essential to trees, and in particular there is a precise notion of \emph{unlabeled tree}.
\begin{definition}
  Let \( x \in \ty \).
  The \DefnFont{structure} of \( x \), \( \tystruct{x} \), is the underlying unlabeled tree of \( x \).
\end{definition}

\begin{example} \label{ex:typestructure}
  Revisiting the type from Example~\ref{ex:typeorder},
  the type's structure is the following unlabeled tree:
  
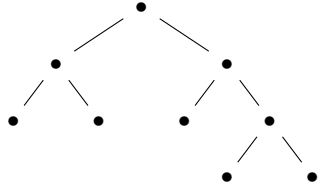
\begin{equation*}
    \begin{tikzcd}[row sep={1em}, column sep={1.6em, between origins}]
      & & & \arrow[dll, dash] \bullet \arrow[drr, dash] & & & & \\
      & \arrow[dl, dash] \bullet \arrow[dr, dash] & & & & \arrow[dl, dash] \bullet \arrow[dr, dash] & & \\
      \bullet & & \bullet & & \bullet & & \arrow[dl, dash] \bullet \arrow[dr, dash] & \\
      & & & & & \bullet & & \bullet
    \end{tikzcd}
  \end{equation*}
\end{example}

Note that the ``underlying unlabeled tree'' of any elementary type would simply be the tree with exactly one vertex, and this means that all elementary types have the same structure.
And clearly, the tree with one vertex is different from the structure of any non-elementary type \( a \tyto b \).

\subsection{Typed linear maps}

The main ingredient of the definition of the \LinearMaps family is the standard construction which makes \( \LinearOps(\MyH_1, \MyH_2) \), the set of all linear maps with domain \( \MyH_1 \) and codomain \( \MyH_2 \), into a Hilbert space for any pair of finite-dimensional Hilbert spaces \( \MyH_1 \) and \( \MyH_2 \).
\( \LinearOps(\MyH_1, \MyH_2) \) is equipped with the Hilbert-Schmidt inner product, which, we recall, is
defined so that, for any pair of linear maps \( f,g \in \LinearOps(\MyH_1, \MyH_2) \),
\begin{equation} \label{eq:hilbertschmidt}
  \braket{f | g}_{\text{HS}} = \Trace{\Adj{f} \circ g},
\end{equation}
where $\circ$ denotes the standard composition of linear maps. Note that \( \Adj{f} \) is defined with respect to the inner products of the Hilbert spaces \( \MyH_1 \) and \( \MyH_2 \).
If \( \MyH_1 \) and \( \MyH_2 \) are both finite-dimensional, then \( \LinearOps(\MyH_1, \MyH_2) \) is finite-dimensional with dimension given by \( {\Dimension \LinearOps(\MyH_1, \MyH_2) = \Dimension \MyH_1 \cdot \Dimension \MyH_2} \).



For each system \( \Sys{A} \in \alphabet \), let \( \HilSpace{\Sys{A}} \) be a finite-dimensional complex Hilbert space chosen to model system \Sys{A}.
Given an elementary type\footnote{We caution the reader against interpreting the elementary type $\Sys{A}_1 \cdots \Sys{A}_n$ as a multiplication or tensor product of quantum states: the notation $\Sys{A}_1 \cdots \Sys{A}_n$ merely represents the label we attach to the composite quantum system made of the systems $\Sys{A}_1$, \ldots, $\Sys{A}_n$.} \( \epsilon = \Sys{A}_1 \cdots \Sys{A}_n \) of length \( n > 0 \), with \( \Sys{A}_j \in \alphabet \) for each \( {j \in \{1,...,n\}} \), we define
\begin{equation}
  \HilSpace{\epsilon} := \bigotimes\limits_{j = 1}^n \HilSpace{\Sys{A}_j} .
\end{equation}




Now we define the \LinearMaps family so that the trivial type \( \Sys{I} \) is associated with the canonical one-dimensional Hilbert space, elementary types are associated with operators on (tensor products of) the base Hilbert spaces, and arrow types are associated with the construction of Hilbert spaces of linear maps.
Note that we write \( \LinearOps( \HilSpace{x} ) \) as an abbreviation for \( \LinearOps( \HilSpace{x} , \HilSpace{x} ) \).
\begin{definition} \label{def:linearmaps}
  
  Given \( x \in \ty \), define a Hilbert space $\LinearMaps(x)$ as follows:
  
  \begin{equation} \label{eq:linearmapsdefinition}
    \LinearMaps(x) =
    \begin{cases*}
      \mathbb{C} & if \( x = \Sys{I} \) \\
      \LinearOps\big( \HilSpace{x} \big) & if \( x = \Sys{A_1} \cdots  \Sys{A}_n  \in \et \setminus \{\Sys{I}\} \) \\
      \LinearOps\big(\LinearMaps(a), \hspace{0.1em} \LinearMaps(b)\big) & if \( x = a \tyto b \)
    \end{cases*}
  \end{equation}
\end{definition}

A couple of observations on the latter definition:
\begin{itemize}
  \item For all non-elementary types $x$, the Hilbert space $\LinearMaps(x)$ is defined with the Hilbert-Schmidt inner product, precisely as in~\eqref{eq:hilbertschmidt}.
  \item Instead of letting \( \LinearMaps(\Sys{I}) \) be \( \LinearOps(\mathbb{C}) \), which would directly match the definition of \LinearMaps for other elementary types, we let it be \( \mathbb{C} \).
This is more convenient, and the overall definition is still consistent with \Sys{I} being an elementary type, because \( \mathbb{C} \) and \( \LinearOps(\mathbb{C}) \) are isomorphic as Hilbert spaces.
\end{itemize}

Quantum objects such as states, channels, and superchannels now fall within the domain encompassed by the \LinearMaps family.
Given some arbitrary base systems \( \Sys{A}, \Sys{B}, \Sys{C}, \Sys{D} \in \alphabet \), by definition we have
\begin{gather}
  \LinearMaps(\Sys{A}) = \LinearOps(\HilSpace{\Sys{A}}) \\
  \LinearMaps(\Sys{A} \tyto \Sys{B}) = \LinearOps\big( \LinearOps(\HilSpace{\Sys{A}}), \hspace{0.15em} \LinearOps(\HilSpace{\Sys{B}}) \big) \\
  \LinearMaps\big( (\Sys{A} \tyto \Sys{B}) \tyto (\Sys{C} \tyto \Sys{D}) \big) = \LinearOps\Big( \LinearOps\big( \LinearOps(\HilSpace{\Sys{A}}), \hspace{0.15em} \LinearOps(\HilSpace{\Sys{B}}) \big) , \hspace{0.15em} \LinearOps\big( \LinearOps(\HilSpace{\Sys{C}}), \hspace{0.15em} \LinearOps(\HilSpace{\Sys{D}}) \big)  \Big)
\end{gather}
Hence by the standard definitions of quantum state, channel, and superchannel, a (mixed) state of system \Sys{A} is an operator \( \rho \in \LinearMaps(\Sys{A}) \);
a channel with input system \Sys{A} and output system \Sys{B} is a linear map \( {\higherfont{M} \in \LinearMaps(\Sys{A} \tyto \Sys{B})} \);
and a superchannel with input a channel of input system \Sys{A} and output system \Sys{B}, and output a channel of input system \Sys{C} and output system \Sys{D}, is a linear map \( {\Theta \in \LinearMaps\big( (\Sys{A} \tyto \Sys{B}) \tyto (\Sys{C} \tyto \Sys{D}) \big)} \).


Of course, states form a strict subset of \( \LinearMaps(\Sys{A}) \), channels a subset of \( \LinearMaps(\Sys{A} \tyto \Sys{B}) \), and superchannels a subset of \( \LinearMaps((\Sys{A} \tyto \Sys{B}) \tyto (\Sys{C} \tyto \Sys{D})) \).
Indeed, one of our goals in this work is to identify within each space \( \LinearMaps(x) \) a subset of physically meaningful maps (see Section~\ref{sec:completelypositivemaps}).
\begin{remark}
		The dimensions of $\LinearMaps(\mathrm{AB})$ and $\LinearMaps(\mathrm{A}\rightarrow \mathrm{B})$ are the same. However, we insist on regarding these two sets as separate and, a priori, inequivalent. This is the key formal move that distinguishes type-based approaches from non-type-based approaches. We insist that linear maps and their Hilbert spaces always be accompanied (we often say ``labeled'') by a specified type. To emphasize this point of view, we often call these objects (spaces of) ``typed linear maps.'' In this setting, two Hilbert spaces with the same dimension but different types cannot be considered equal (or identified). We do this because we want to maintain a crisp correspondence between the linear map as a mathematical object and what it represents physically. By looking at the type of a typed linear map, we can immediately understand what kind of physical object (or process) it describes.
\end{remark}

Following the convention in type theory, we use a colon as notation to indicate the type of a linear map.
\begin{notation}
  We write the statement \( \higherfont{M} \in \LinearMaps(x) \) equivalently as
  \begin{equation}
    \higherfont{M} : x
  \end{equation}
\end{notation}
Notice that in the context of typed linear maps, two maps are the same \emph{only if} they are of the same type. If this condition is not satisfied, two typed linear maps cannot be equal, even when they have ``equivalent'' functional action. We also caution the reader not to understand $\higherfont{M} : x$ as $\higherfont{M}$ acting on $x$. Here, $x$ expresses not just the information of what kind of input \( \higherfont{M} \) accepts, but also what kind of output \(\higherfont{M} \) returns.
Instead, it would be correct to view $\higherfont{M}$ as acting on maps of its input type, i.e.\ elements of $\LinearMaps(\tyin{x})$.

Composition makes sense for our typed linear maps.
Given maps \( \higherfont{M} : a \tyto b \) and \( \higherfont{N} : b \tyto c \), by definition we have
\begin{equation}
  \higherfont{M} \in \LinearOps\big(\LinearMaps(a), \LinearMaps(b)\big) ,
\end{equation}
and
\begin{equation}
  \higherfont{N} \in \LinearOps\big(\LinearMaps(b), \LinearMaps(c)\big) .
\end{equation}
Thus \( \higherfont{M} \) and \( \higherfont{N} \) are composable as functions, and since the composite of linear maps is again linear, we have
\begin{equation}
  \higherfont{N} \circ \higherfont{M} \in \LinearOps\big(\LinearMaps(a), \LinearMaps(c)\big) .
\end{equation}
In other words, we have \( \higherfont{N} \circ \higherfont{M} : a \tyto c \), just as one would want.

\section{The parallel product} \label{sec:parallelproduct}

In this section, we seek to construct an operation on typed linear maps, to be denoted \parmap, suitable for taking any two maps \( \higherfont{M} : x \) and \( \higherfont{N} : y \) and producing a map \( \higherfont{M} \parmap \higherfont{N} : x \partype y \) to be called the \emph{parallel product} of \( \higherfont{M} \) and \( \higherfont{N} \).
The map \( \higherfont{M} \parmap \higherfont{N} \) will express the idea that these two linear maps are applied in parallel in an operational scenario.
We must define an operation on two different levels:
we must define a type \( x \partype y \), and we must define a map \( \higherfont{M} \parmap \higherfont{N} \) of type \( x \partype y \).

The parallel product operation should work in a reasonable way for maps of any types, even when the types have different orders, or more generally, different structures.
We have chosen the term ``parallel product'' because it is descriptive of the role that the tensor product operation plays in ordinary quantum theory.
However, our starting point is the observation that the tensor product is \emph{not} suitable for higher-order maps of arbitrary types \( x \) and \( y \).

\subsection{The tensor product as parallel product?} \label{sec:tensorproductdiscussion}


Let us recall the usual role of the tensor product.
For a pair of systems \Sys{A} and \Sys{B}, modeled by Hilbert spaces \( \HilSpace{\Sys{A}} \) and \( \HilSpace{\Sys{B}} \), the composite system \Sys{AB} is modeled by the tensor product space \( \HilSpace{\Sys{A}} \otimes \HilSpace{\Sys{B}} \).
This is the role of the tensor product of spaces.
Given a state \( \rho \) of system \Sys{A} and a state \( \sigma \) of system \Sys{B}, the tensor product state \( \rho \otimes \sigma \) is a fully unentangled state of the composite system \Sys{AB}.
This state asserts that the condition of system \Sys{AB} is that the \Sys{A} subsystem is prepared in state \( \rho \) while independently the \Sys{B} subsystem is prepared in state \( \sigma \).
We might say of this situation that the two states are prepared simultaneously, in parallel.

The tensor product has yet another role in the way it operates on quantum transformations.
Recall that for any two linear maps \( {f \in \LinearOps(\MyH_1, \MyH_2)} \) and \( {g \in \LinearOps(\MyH_3, \MyH_4)} \), their tensor product is a linear map
\begin{equation} \label{eq:tensorproducttype}
  f \otimes g \in \LinearOps(\MyH_1\otimes \MyH_3, \hspace{0.15em} \MyH_2 \otimes \MyH_4) .
\end{equation}
Moreover, \( f \otimes g \) may be defined by the following formula:
\begin{equation} \label{eq:tensorproductformula}
  \forall u \in \MyH_1 , \forall v \in \MyH_3 , \ (f \otimes g)(u \otimes v) = f(u) \otimes g(v) .
\end{equation}
This formula extends to a unique linear map on the domain \( \MyH_1 \otimes \MyH_3 \), even though it specifies the action only for inputs in the form of a tensor product, because the domain is spanned by these elements.

Let \( \Sys{A}, \Sys{B}, \Sys{C}, \Sys{D} \) be systems, and suppose we have two channels:
\begin{equation}
  \begin{split}
    \higherfont{M} \in \LinearOps\big( \LinearOps(\HilSpace{\Sys{A}} ) , \hspace{0.15em} \LinearOps(\HilSpace{\Sys{B}} ) \big) \\
    \higherfont{N} \in \LinearOps\big( \LinearOps(\HilSpace{\Sys{C}} ) , \hspace{0.15em} \LinearOps(\HilSpace{\Sys{D}} ) \big)
  \end{split}
\end{equation}
Then, in accordance with~\eqref{eq:tensorproducttype}, the tensor product of \( \higherfont{M} \) with \( \higherfont{N} \) is a linear map
\begin{equation} \label{eq:channeltensorproduct}
  \higherfont{M} \otimes \higherfont{N} \in
  \LinearOps\big( \LinearOps(\HilSpace{\Sys{A}} ) \otimes \LinearOps(\HilSpace{\Sys{C}} ) , \ \LinearOps(\HilSpace{\Sys{B}} ) \otimes \LinearOps(\HilSpace{\Sys{D}} ) \big) .
\end{equation}
Now, by the properties of finite-dimensional complex Hilbert spaces, for any finite-dimensional \( \MyH_1,\MyH_2,\MyH_3,\MyH_4 \) we have
\begin{equation} \label{eq:mapspacetensorspaceidentification}
  \LinearOps(\MyH_1 , \MyH_2) \otimes \LinearOps(\MyH_3, \MyH_4)
  \ = \
  \LinearOps(\MyH_1 \otimes \MyH_3, \MyH_2\otimes \MyH_4)
\end{equation}
Therefore, we can rewrite~\eqref{eq:channeltensorproduct} as follows:
\begin{equation} \label{eq:tensorproductofchannels}
  \higherfont{M} \otimes \higherfont{N} \in
  \LinearOps\big( \LinearOps(\HilSpace{\Sys{A}} \otimes \HilSpace{\Sys{C}} ) , \ \LinearOps(\HilSpace{\Sys{B}} \otimes \HilSpace{\Sys{D}} ) \big) .
\end{equation}
Thus, the tensor product channel operates on input states of the composite system \Sys{AC} and returns output states of the composite system \Sys{BD}.
In other words, \( \higherfont{M} \otimes \higherfont{N} \) is of type \( \Sys{AC} \tyto \Sys{BD} \).
And in accordance with~\eqref{eq:tensorproductformula}, any product state \( \rho \otimes \sigma \) of system \Sys{AC} is sent to the product state \( \higherfont{M}(\rho) \otimes \higherfont{N}(\sigma) \).
We might say that in the action of \( \higherfont{M} \otimes \higherfont{N} \), the two channels act in parallel, with \( \higherfont{M} \) acting on the \Sys{A} subsystem and \( \higherfont{N} \) acting on the \Sys{C} subsystem.



Now having seen that the tensor product operation provides something like a parallel product of objects both at the level of states and that of channels,
and having seen that in both cases this notion of parallel product is suitably related to the notion of composite system,
we must observe that the tensor product gives an inadequate result if we try using it to combine a state with a channel.
We emphasize that there is nothing \emph{mathematically} wrong, rather it is a matter of inadequacy with respect to the way we use the mathematics for the operational description of quantum theory.

Suppose we have a state \( \rho \) of system \Sys{A} and a channel \( \higherfont{M} \) from system \Sys{C} to system \Sys{D}:
\begin{align}
  & \rho \in \LinearOps(\HilSpace{\Sys{A}})
  && \higherfont{M} \in \LinearOps\big(\LinearOps(\HilSpace{\Sys{C}}), \hspace{0.15em} \LinearOps(\HilSpace{\Sys{D}})\big)
\end{align}
In accordance with~\eqref{eq:tensorproducttype}, the tensor product of \( \rho \) and \( \higherfont{M} \) is a linear map
\begin{equation} \label{eq:parallelproductmotivation1}
  \rho \otimes \higherfont{M} \in
  \LinearOps\big( \HilSpace{\Sys{A}} \otimes \LinearOps(\HilSpace{\Sys{C}}) , \hspace{0.15em} \HilSpace{\Sys{A}} \otimes \LinearOps(\HilSpace{\Sys{D}}) \big) .
\end{equation}
Now something has gone wrong, because neither of the Hilbert spaces \( {\HilSpace{\Sys{A}} \otimes \LinearOps(\HilSpace{\Sys{C}})} \) or \( {\HilSpace{\Sys{A}} \otimes \LinearOps(\HilSpace{\Sys{D}})} \) has a clear operational meaning for the way we model quantum theory.
\HilSpace{\Sys{A}} is a Hilbert space we have chosen to model system \Sys{A}, but we do that by defining states of system \Sys{A} to be a subset of the space of linear operators \( \LinearOps(\HilSpace{\Sys{A}}) \).
Even if we decided to view a (non-zero) \( v \in \HilSpace{\Sys{A}} \) as a representative of a pure state of system \Sys{A}, there is no operational meaning to the output \( \rho(v) \otimes \higherfont{M}(\sigma) \) that results from applying \( \rho \otimes \higherfont{M} \) to a product input \( v \otimes \sigma \), where \( \sigma \) is a state of system \Sys{C}.
Ultimately, there is no kind of quantum object that \( \rho \otimes \higherfont{M} \) could be understood as modeling.

The map \( \rho \otimes \higherfont{M} \) may be unsuitable for our purposes, but there is another way to use the tensor product with \( \rho \) and \( \higherfont{M} \) such that the result is always a legitimate channel.
This is the channel that will be formally defined as \( \rho \parmap \higherfont{M} \) when the full \parmap operation is defined later in this section.
The idea of \( \rho \parmap \higherfont{M} \) is simple: since \( \higherfont{M} \) takes a state as input and returns a state as output, and since \( \rho \) takes no input, we can define \( \rho \parmap \higherfont{M} \) as a map that takes an input \( \sigma \in \LinearOps(\HilSpace{\Sys{C}}) \) and passes it to \( \higherfont{M} \), appends the resulting output \( \higherfont{M}(\sigma) \) to \( \rho \) using the tensor product, and therefore returns \( \rho \otimes \higherfont{M}(\sigma) \) as the final output.
It is straightforward to prove that this results in a legitimate linear map (moreover, a channel)
\begin{equation} \label{eq:typeofparallelproductofstateandchannel}
  \rho \parmap \higherfont{M} \in \LinearOps\big(\LinearOps(\HilSpace{\Sys{C}}), \hspace{0.15em} \LinearOps(\HilSpace{\Sys{A}} \otimes \HilSpace{\Sys{D}})\big)
\end{equation}
which acts according to the formula
\begin{equation} \label{eq:statechannelparallel}
  \forall \sigma \in \LinearOps(\HilSpace{\Sys{C}}) , \ (\rho \parmap \higherfont{M})(\sigma) = \rho \otimes \higherfont{M}(\sigma) .
\end{equation}
Note that according to~\eqref{eq:typeofparallelproductofstateandchannel}, \( \rho \parmap \higherfont{M} \) is of type \( \Sys{C} \tyto \Sys{AD} \).

\subsection{Defining the generalized parallel product}

The latter considerations lead us to the general definition of the parallel product of typed linear maps.
The basic form of the parallel product is as follows:
given typed linear maps \( \higherfont{M} \in \LinearMaps(x) \) and \( \higherfont{N} \in \LinearMaps(y) \),
their parallel product is a typed linear map denoted \( \higherfont{M} \parmap \higherfont{N} \).
The type of \( \higherfont{M} \parmap \higherfont{N} \) depends only on the types \( x \) and \( y \), not the particular maps.
We therefore introduce another operation, the parallel product of types, so that the product of \( x \) and \( y \), denoted \( x \partype y \), is exactly the type of \( \higherfont{M} \parmap \higherfont{N} \).
In other words, the definition of the parallel product of types is chosen to ensure that
\begin{equation}
  \higherfont{M} \parmap \higherfont{N} \in \LinearMaps(x \partype y) .
\end{equation}

We will begin by defining the parallel product of maps and observing how the definition captures the main insights of the previous section.
We will then state the corresponding definition of the parallel product of types.
Then, after several explanatory comments about the definitions, we will proceed to provide examples of how these definitions are put to use.

\begin{definition}[Parallel product map] \label{def:parallelmap}
  Let \( \higherfont{M} : x \) and \( \higherfont{N} : y \) for \( x,y\in\ty \).
  The linear map \( \higherfont{M} \parmap \higherfont{N} \) is defined according to the following formulae:
  for all \( \rho : a \) and for all \( \sigma : c \)
  \begin{align} \label{eq:parallelmapdefinition}
    & \higherfont{M} \parmap \higherfont{N} := \notag \\
	& \begin{cases*}
		\higherfont{M} \otimes \higherfont{N} & if \( x,y\in\et \) \\[0.25em]
		(\higherfont{M} \parmap \higherfont{N})(\rho \parmap \sigma) = \higherfont{M}(\rho) \parmap \higherfont{N}(\sigma) & if \( x = a \tyto b \), \( y = c \tyto d \), and \( \tyord{x} = \tyord{y} \) \\[0.25em]
		(\higherfont{M} \parmap \higherfont{N})(\sigma) = \higherfont{M} \parmap \higherfont{N}(\sigma) & if \( y = c \tyto d \) and \( \tyord{x} < \tyord{y} \) \\[0.25em]
		(\higherfont{M} \parmap \higherfont{N})(\rho) = \higherfont{M}(\rho) \parmap \higherfont{N} & if \( x = a \tyto b \) and \( \tyord{x} > \tyord{y} \)
	\end{cases*}
  \end{align}
\end{definition}
We observe how the four cases of the definition capture different insights from the previous section:
\begin{enumerate}
  \item The definition's first case says that the parallel product of elementary maps (i.e.\ quantum states) is just their tensor product. Thus, in the case of elementary maps we not only imitate the tensor product, but reproduce it exactly.
  \item The second case applies when \( \tyord{x} = \tyord{y} \), and it takes exactly the form of the definition of the tensor product that we saw in the last section in Eq.~\eqref{eq:tensorproductformula}, except that the tensor product operation is everywhere replaced by the parallel product operation.
  This means that for channels the parallel product not only imitates, but also reproduces the tensor product exactly: \( \higherfont{M} \parmap \higherfont{N}(\rho \parmap \sigma) \) is \( \higherfont{M}(\rho) \parmap \higherfont{N}(\sigma) \), but the latter is by definition the same as \( \higherfont{M}(\rho) \otimes \higherfont{N}(\sigma) \) because \( \higherfont{M}(\rho) \) and \( \higherfont{N}(\sigma) \) are maps of order zero.
  For orders greater than one, the parallel product formula imitates the tensor product formula for channels, but they are not generally identical.
  \item The third and fourth cases are defined symmetrically for when \( \tyord{x} < \tyord{y} \), and when \( \tyord{x} > \tyord{y} \), respectively.
  The key idea is to imitate the solution to the problem of the tensor product that we demonstrated at the end of the previous section.
  The formula defining the parallel product here is exactly like Eq.~\eqref{eq:statechannelparallel}, except (again) the tensor product operation is everywhere replaced by the parallel product operation.
\end{enumerate}

Definition~\ref{def:parallelmap} ensures that both the input and output spaces of the linear map \( \higherfont{M} \parmap \higherfont{N} \) are certain spaces of typed linear maps,
which means not only that \( \higherfont{M} \parmap \higherfont{N} \) may be assigned a type in our type system, but that Definition~\ref{def:parallelmap} already forces a particular choice of type.
Thus, the motivation we cited for the definition of the parallel product of maps also leads us to adopt the following definition of the parallel product of types.
\begin{definition}[Parallel product type] \label{def:paralleltype}
  Let \( x,y\in\ty \).
  The type \( x \partype y \in \ty \) is defined as:
  \begin{equation}
    x \partype y =
    \begin{cases*}
      \Sys{A}_1\cdots \Sys{A}_j \Sys{B}_1 \cdots \Sys{B}_k & if \( x = \Sys{A}_1\cdots \Sys{A}_j \) and \( y = \Sys{B}_1 \cdots \Sys{B}_k \) \\[0.25em]
      a \partype c \tyto b \partype d & if \( x = a \tyto b \), \( y = c \tyto d \), and \( \tyord{x} = \tyord{y} \) \\[0.25em]
      c \tyto x \partype d & if \( y = c \tyto d \) and \( \tyord{x} < \tyord{y} \) \\[0.25em]
      a \tyto b \partype y & if \( x = a \tyto b \) and \( \tyord{x} > \tyord{y} \)
    \end{cases*}
  \end{equation}
\end{definition}
Note that we write \( \Sys{A}_1\cdots \Sys{A}_j \Sys{B}_1 \cdots \Sys{B}_k \) to denote the concatenation of the sequences \( \Sys{A}_1\cdots \Sys{A}_j \) and \( \Sys{B}_1 \cdots \Sys{B}_k \), which means in particular that \( \Sys{I}\Sys{A} = \Sys{A} = \Sys{A}\Sys{I} \), as \Sys{I} denotes the empty sequence.

Consider the elementary case of Definition~\ref{def:paralleltype}.
To see that this matches the definition of the parallel product of maps, one needs only to compare the form of the usual tensor product (cf.\ Eq.~\eqref{eq:tensorproducttype}), with Eq.~\eqref{eq:mapspacetensorspaceidentification}.
In each of the non-elementary cases, the form of the parallel product of types merely records the input and output spaces belonging to the parallel product map that are implicitly specified in Definition~\ref{def:parallelmap}.

Now we pause to discuss the inductive nature of the latter definitions.
The inductive form of Definition~\ref{def:paralleltype} is such that each time an expression \( w \partype z \) is used therein to help define \( x \partype y \), the types \( w \) and \( z \) are both of lower order than the maximum order of \( x \) and \( y \).
Moreover, we have a valid base case for when \( \tyord{x} = \tyord{y} = 0 \).
Thus, the definition works by induction on the number \( \max\{\tyord{x}, \tyord{y}\} \).

Definition~\ref{def:parallelmap} is also fully inductive, as the parallel product of $ \higherfont{M}$ and $ \higherfont{N}$ is defined via the parallel product of objects of lower order, such as their inputs $\rho$ and $\sigma$. Moreover, it is non-trivial, but we are able prove that this specification gives a well-defined linear map \( \higherfont{M} \parmap \higherfont{N} : x \partype y \) in all cases, and that $\parmap$ is a bilinear product.
The main reason for the non-triviality is the case where \( x \) and \( y \) are non-elementary of equal order.
To produce a well-defined function, we must take the linear extension of the given formula to the full domain \( \LinearMaps(a \partype c) \).
What must be shown then is that the subset of elements of the form \( \rho \parmap \sigma \) is a spanning subset of \( \LinearMaps(a \partype c) \).
This is indeed the case, as we demonstrate in our full results \cite{steakleyOperationalHigherorderQuantum,steakleyTypebasedFrameworkHigherorder,zanoniMapsHigherOrderMaps2025}.

In the following example, we apply Definition~\ref{def:paralleltype} step-by-step and confirm that it produces the desired result.%

\begin{example} \label{ex:partypeexample1}
  Let \( x = \Sys{A} \tyto \Sys{B} \) and \( y = \Sys{C} \tyto \Sys{D} \).
  \( x \) and \( y \) are both of order one, and so according to~\eqref{eq:parallelmapdefinition} we must use the symmetric case of the parallel product:
  \begin{equation}
    x \partype y \enspace =
    \begin{tikzcd}[row sep={1em}, column sep={1.3em, between origins}]
      & \tyto & \\
      \Sys{A} \arrow[from=ur, dash] & & \Sys{B} \arrow[from=ul, dash]
    \end{tikzcd}
    \ \partype \
    \begin{tikzcd}[row sep={1em}, column sep={1.3em, between origins}]
      & \tyto & \\
      \Sys{C} \arrow[from=ur, dash] & & \Sys{D} \arrow[from=ul, dash]
    \end{tikzcd}
    \enspace = \enspace
    \begin{tikzcd}[row sep={1em}, column sep={1.7em, between origins}]
      & \tyto & \\
      \Sys{A} \partype \Sys{C} \arrow[from=ur, dash] & & \Sys{B} \partype \Sys{D} \arrow[from=ul, dash]
    \end{tikzcd}
  \end{equation}
  For elementary types, \partype is computed by concatenation, and so we have \( \Sys{A} \partype \Sys{C} = \Sys{AC} \) and \( \Sys{B} \partype \Sys{D} = \Sys{BD} \).
  Therefore, the final result is
  \begin{equation}
    x \partype y \enspace =
    \begin{tikzcd}[row sep={1em}, column sep={1.3em, between origins}]
      & \tyto & \\
      \Sys{AC} \arrow[from=ur, dash] & & \Sys{BD} \arrow[from=ul, dash]
    \end{tikzcd}
  \end{equation}
    Since \( x \) and \( y \) are both the sort of type that would describe a channel, what we want in this case is to reproduce the behavior of the ordinary tensor product.
  And indeed, the type \( \Sys{AC} \tyto \Sys{BD} \) is exactly what we obtained in the previous section, cf.\ Eq.~\eqref{eq:tensorproductofchannels} and the surrounding discussion.
  
\end{example}

In general, to apply Definition~\ref{def:paralleltype} to compute \( x \partype y \) involves several steps: first compare the orders of \( x \) and \( y \), and depending on the result we recurse (compute \partype again) onto subtrees of \( x \) and \( y \).
Here we see an example where computing \( x \partype y \) involves alternating invocations of both the symmetric and asymmetric cases.
\begin{example} \label{ex:partypeexample2}
  Let \( x = \Sys{A} \tyto (\Sys{B}\tyto \Sys{C}) \) and \( y = (\Sys{D}\tyto \Sys{E}) \tyto \Sys{F} \).
  Note that \( \tystruct{x} \neq \tystruct{y} \), that is,
  \begin{equation}
    \begin{tikzcd}[row sep={1em}, column sep={1.3em, between origins}]
      & \bullet & & \\
      \bullet \arrow[from=ur, dash] & & \bullet \arrow[from=ul, dash] & \\
      & \bullet \arrow[from=ur, dash] & & \bullet \arrow[from=ul, dash]
    \end{tikzcd}
    \enspace \neq \enspace
    \begin{tikzcd}[row sep={1em}, column sep={1.3em, between origins}]
      & & \bullet & \\
      & \bullet \arrow[from=ur, dash] & & \bullet \arrow[from=ul, dash] \\
      \bullet \arrow[from=ur, dash] & & \bullet \arrow[from=ul, dash] &
    \end{tikzcd}
    \ .
  \end{equation}
  \( x \) and \( y \) are both of order two, so the first step is symmetric:
  \begin{equation} \label{eq:partypeexample2_firststep}
    \begin{tikzcd}[row sep={1em}, column sep={1.3em, between origins}]
      & \tyto & & \\
      \Sys{A} \arrow[from=ur, dash] & & \tyto \arrow[from=ul, dash] & \\
      & \Sys{B} \arrow[from=ur, dash] & & \Sys{C} \arrow[from=ul, dash]
    \end{tikzcd}
    \enspace \partype \enspace
    \begin{tikzcd}[row sep={1em}, column sep={1.3em, between origins}]
      & & \tyto & \\
      & \tyto \arrow[from=ur, dash] & & \Sys{F} \arrow[from=ul, dash] \\
      \Sys{D} \arrow[from=ur, dash] & & \Sys{E} \arrow[from=ul, dash] &
    \end{tikzcd}
    \quad = \quad
    \begin{tikzcd}[row sep={1em}, column sep={3.6em, between origins}]
      & \arrow[dl, dash] \tyto \arrow[dr, dash] & \\
      \Sys{A} \partype (\Sys{D} \tyto \Sys{E}) & & (\Sys{B} \tyto \Sys{C}) \partype \Sys{F}
    \end{tikzcd}
  \end{equation}
  Then we compute the recursive applications of \partype demanded by the right-hand side of~\eqref{eq:partypeexample2_firststep}:
  \begin{equation} \label{eq:partypeexample2_secondstep}
    \Sys{A}
    \enspace \partype \enspace
    \begin{tikzcd}[row sep={1em}, column sep={1.3em, between origins}]
      & \arrow[dl, dash] \tyto \arrow[dr, dash] & \\
      \Sys{D} & & \Sys{E}
    \end{tikzcd}
    \quad = \quad
    \begin{tikzcd}[row sep={1em}, column sep={1.6em, between origins}]
      & \arrow[dl, dash] \tyto \arrow[dr, dash] & \\
      \Sys{D} & & \Sys{A} \partype \Sys{E}
    \end{tikzcd}
    \quad = \quad
    \begin{tikzcd}[row sep={1em}, column sep={1.3em, between origins}]
      & \arrow[dl, dash] \tyto \arrow[dr, dash] & \\
      \Sys{D} & & \Sys{AE}
    \end{tikzcd}
  \end{equation}
  \begin{equation} \label{eq:partypeexample2_thirdstep}
    \begin{tikzcd}[row sep={1em}, column sep={1.3em, between origins}]
      & \arrow[dl, dash] \tyto \arrow[dr, dash] & \\
      \Sys{B} & & \Sys{C}
    \end{tikzcd}
    \enspace \partype \enspace
    \Sys{F}
    \quad = \quad
    \begin{tikzcd}[row sep={1em}, column sep={1.6em, between origins}]
      & \arrow[dl, dash] \tyto \arrow[dr, dash] & \\
      \Sys{B} & & \Sys{C} \partype \Sys{F}
    \end{tikzcd}
    \quad = \quad
    \begin{tikzcd}[row sep={1em}, column sep={1.3em, between origins}]
      & \arrow[dl, dash] \tyto \arrow[dr, dash] & \\
      \Sys{B} & & \Sys{CF}
    \end{tikzcd}
  \end{equation}
  Plugging these results back into~\eqref{eq:partypeexample2_firststep}, the final result is:
  \begin{equation} \label{eq:partypeexample2_finalresult}
    x \partype y
    \quad = \quad
    \begin{tikzcd}[row sep={1em}, column sep={1.3em, between origins}]
      & & & \tyto & & & \\
      & \tyto \arrow[from=urr, dash] & & & & \tyto \arrow[from=ull, dash] & \\
      \Sys{D} \arrow[from=ur, dash] & & \Sys{AE} \arrow[from=ul, dash] & & \Sys{B} \arrow[from=ur, dash] & & \Sys{CF} \arrow[from=ul, dash]
    \end{tikzcd}
  \end{equation}
\end{example}

Finally, we introduce by example an alternative way to compute the parallel product type (for more details, see \cite{steakleyOperationalHigherorderQuantum,steakleyTypebasedFrameworkHigherorder,zanoniMapsHigherOrderMaps2025}), by splitting the process into two separate phases.
We revisit the types of Example~\ref{ex:partypeexample2}.
\begin{example} \label{ex:partypeexample3}
  Let \( x = \Sys{A} \tyto (\Sys{B}\tyto \Sys{C}) \) and \( y = (\Sys{D}\tyto \Sys{E}) \tyto \Sys{F} \).
  Graphically, we have:
  \begin{align}
    x
    \ = \
    \begin{tikzcd}[row sep={1em}, column sep={1.3em, between origins}, ampersand replacement=\&]
      \& \tyto \& \& \\
      \Sys{A} \arrow[from=ur, dash] \& \& \tyto \arrow[from=ul, dash] \& \\
      \& \Sys{B} \arrow[from=ur, dash] \& \& \Sys{C} \arrow[from=ul, dash]
    \end{tikzcd}
    &&
    y
    \ = \
    \begin{tikzcd}[row sep={1em}, column sep={1.3em, between origins}, ampersand replacement=\&]
      \& \& \tyto \& \\
      \& \tyto \arrow[from=ur, dash] \& \& \Sys{F} \arrow[from=ul, dash] \\
      \Sys{D} \arrow[from=ur, dash] \& \& \Sys{E} \arrow[from=ul, dash] \&
    \end{tikzcd}
  \end{align}
  The first phase consists in modifying \( x \) and \( y \), by augmenting their trees with new vertices labeled by the trivial type \Sys{I} (each one placed under a new arrow-labeled vertex),
  until
  (i) the modified forms of \( x \) and \( y \) are of equal order, and (ii) every pair of matching subtrees, of the modified forms of \( x \) and \( y \), are of equal order.
  In addition, we must only add new \Sys{I}-labeled vertices \emph{on the left}.

  Now, proceeding with \( x \) and \( y \).
  \( x \) and \( y \) are the same order, and thus we need not change anything at this level.
  So, we pass to the subtrees of \( x \) and \( y \).
  First the left subtrees: we compare \( \Sys{A} \) with \( \Sys{D} \tyto \Sys{E} \) and see that \Sys{A} is of lesser order.
  So, we change \( \Sys{A} \) into \( \Sys{I} \tyto \Sys{A} \) within \( x \):
  \begin{equation} \label{eq:partypeexample3_1}
    \begin{tikzcd}[row sep={1em}, column sep={1.3em, between origins}]
      & \tyto & & \\
      \Sys{A} \arrow[from=ur, dash] & & \tyto \arrow[from=ul, dash] & \\
      & \Sys{B} \arrow[from=ur, dash] & & \Sys{C} \arrow[from=ul, dash]
    \end{tikzcd}
    \enspace \rightsquigarrow \enspace
    \begin{tikzcd}[row sep={1em}, column sep={1.3em, between origins}]
      & & & \tyto & & & \\
      & \tyto \arrow[from=urr, dash] & & & & \tyto \arrow[from=ull, dash] & \\
      \Sys{I} \arrow[from=ur, dash] & & \Sys{A} \arrow[from=ul, dash] & & \Sys{B} \arrow[from=ur, dash] & & \Sys{C} \arrow[from=ul, dash]
    \end{tikzcd}
  \end{equation}
  Now the right subtrees of \( x \) and \( y \): we compare \( \Sys{B} \tyto \Sys{C} \) with \( \Sys{F} \) and see that \Sys{F} is of lesser order. So, we change \( \Sys{F} \) into \( \Sys{I} \tyto \Sys{F} \):
  \begin{equation} \label{eq:partypeexample3_2}
    \begin{tikzcd}[row sep={1em}, column sep={1.3em, between origins}]
      & & \tyto & \\
      & \tyto \arrow[from=ur, dash] & & \Sys{F} \arrow[from=ul, dash] \\
      \Sys{D} \arrow[from=ur, dash] & & \Sys{E} \arrow[from=ul, dash] &
    \end{tikzcd}
    \enspace \rightsquigarrow \enspace
    \begin{tikzcd}[row sep={1em}, column sep={1.3em, between origins}]
      & & & \tyto & & & \\
      & \tyto \arrow[from=urr, dash] & & & & \tyto \arrow[from=ull, dash] & \\
      \Sys{D} \arrow[from=ur, dash] & & \Sys{E} \arrow[from=ul, dash] & & \Sys{I} \arrow[from=ur, dash] & & \Sys{F} \arrow[from=ul, dash]
    \end{tikzcd}
  \end{equation}
  Now, on the right-hand sides of~\eqref{eq:partypeexample3_1} and~\eqref{eq:partypeexample3_2}, we see modified forms of \( x \) and \( y \) that meet the requirements (i) and (ii) described above.
  Let us call these \( x' \) and \( y' \), so that \( {x' = (\Sys{I} \tyto \Sys{A}) \tyto (\Sys{B} \tyto \Sys{C})} \) and \( y' = (\Sys{D} \tyto \Sys{E}) \tyto (\Sys{I} \tyto \Sys{F}) \).
  In fact, there is another way to verify that (i) and (ii) are satisfied, namely by the fact that \( x' \) and \( y' \) have the same structure (i.e.\ the same tree shape).
  
  Now that we have the correct trees \( x' \) and \( y' \), the final step is simply to compute the parallel product type \( x' \partype y' \).
  However, computing the parallel product type is especially simple for a pair of types that have the same structure.
  One may compute \( x' \partype y' \) simply by overlaying the two trees, one over the other, and combining the matching labels together:
  \begin{align}
    \begin{tikzcd}[row sep={1em}, column sep={1em, between origins}, ampersand replacement=\&]
      \& \& \& \tyto \& \& \& \\
      \& \tyto \arrow[from=urr, dash] \& \& \& \& \tyto \arrow[from=ull, dash] \& \\
      \Sys{I} \arrow[from=ur, dash] \& \& \Sys{A} \arrow[from=ul, dash] \& \& \Sys{B} \arrow[from=ur, dash] \& \& \Sys{C} \arrow[from=ul, dash]
    \end{tikzcd}
    \ \partype \
    \begin{tikzcd}[row sep={1em}, column sep={1em, between origins}, ampersand replacement=\&]
      \& \& \& \tyto \& \& \& \\
      \& \tyto \arrow[from=urr, dash] \& \& \& \& \tyto \arrow[from=ull, dash] \& \\
      \Sys{D} \arrow[from=ur, dash] \& \& \Sys{E} \arrow[from=ul, dash] \& \& \Sys{I} \arrow[from=ur, dash] \& \& \Sys{F} \arrow[from=ul, dash]
    \end{tikzcd}
    \ & = \
    \begin{tikzcd}[row sep={1em}, column sep={1.6em, between origins}, ampersand replacement=\&]
      \& \& \& \tyto \& \& \& \\
      \& \tyto \arrow[from=urr, dash] \& \& \& \& \tyto \arrow[from=ull, dash] \& \\
      \Sys{I} \partype \Sys{D} \arrow[from=ur, dash] \& \& \Sys{A} \partype \Sys{E} \arrow[from=ul, dash] \& \& \Sys{B} \partype \Sys{I} \arrow[from=ur, dash] \& \& \Sys{C} \partype \Sys{F} \arrow[from=ul, dash]
    \end{tikzcd} \\
    & = \ \label{eq:partypeexample3_3}
    \begin{tikzcd}[row sep={1em}, column sep={1.6em, between origins}, ampersand replacement=\&]
      \& \& \& \tyto \& \& \& \\
      \& \tyto \arrow[from=urr, dash] \& \& \& \& \tyto \arrow[from=ull, dash] \& \\
      \Sys{D} \arrow[from=ur, dash] \& \& \Sys{AE} \arrow[from=ul, dash] \& \& \Sys{B} \arrow[from=ur, dash] \& \& \Sys{CF} \arrow[from=ul, dash]
    \end{tikzcd}
  \end{align}
  As we can see, the final result on the right-hand side of~\eqref{eq:partypeexample3_3} is exactly the same as the final result of Example~\ref{ex:partypeexample2}, which is shown on the right-hand side of~\eqref{eq:partypeexample2_finalresult}.
\end{example}
Going beyond basic examples, there are many general facts about parallel product types that we are able to prove, as detailed in \cite{steakleyOperationalHigherorderQuantum,steakleyTypebasedFrameworkHigherorder,zanoniMapsHigherOrderMaps2025}.
	An instance is the following fact, which we will use in the next section, and for which a proof is given in the appendix.
	\begin{lemma} \label{lem:parallelproductorder}
		For all $x,y \in \ty$, \begin{equation}\label{eq:paralleltypeorder}\tyord{x \partype y} = \max\{ \tyord{x}, \tyord{y} \}.\end{equation}
	\end{lemma}

\begin{remark}[Comparison with prior work]
  As we noted in the introduction, the parallel product is a generalization of the notion of ``extended event'' defined in \cite[Definition~4.3]{bisioTheoreticalFrameworkHigherorder2019}.
  Using extension as they define it,
  one arbitrary type \( x \in \ty \) can be combined with one elementary type \( E \in \alphabet \).
  In the elementary case, \( x = \Sys{A}_1 \cdots \Sys{A}_n \), the extension, denoted \( x \parallel E \), is computed by concatenation just like the parallel product: \( x \parallel E = \Sys{A}_1 \cdots \Sys{A}_n E \).
  In the non-elementary case, \( x = a \tyto b \), the extension is defined exactly how we define the parallel product: \( x \parallel E = a \tyto b \parallel E \).
  Thus, by the definitions presented in this section, we extend this operation on types to pairs of arbitrary types,
  and we also provide a corresponding operation for arbitrary maps.
\end{remark}

\section{Generalizing completely positive maps} \label{sec:completelypositivemaps}


The main task of this section is to define the \( \KMaps \) family of convex cones, whose purpose is to generalize the established notion of complete positivity to higher-order linear maps of all types in our framework.
The key ingredient of the construction is to define \( \KMaps(a \tyto b) \) inductively as the subset of ``completely \mbox{\KMaps-preserving}'' maps within \( \LinearMaps(a \tyto b) \).
Notably, the definition of the \( \KMaps \) family is stated in terms of the parallel product operation of Section~\ref{sec:parallelproduct}.

\subsection{Defining the \KMaps family}

In quantum theory, complete positivity arises as one of the properties that characterize channels as a subset of the linear maps \( \LinearOps\big(\LinearOps(\HilSpace{\Sys{A}}), \LinearOps(\HilSpace{\Sys{B}})\big) \), for some given input and output systems \Sys{A} and \Sys{B}, respectively.
Channels are required to send states to states, and this entails that they must send positive semidefinite input operators to positive semidefinite output operators.
The reason channels are required to be \emph{completely} positive is that a map \( {\higherfont{M} \in \LinearOps\big(\LinearOps(\HilSpace{\Sys{A}}), \LinearOps(\HilSpace{\Sys{B}})\big)} \) is \emph{not} completely positive precisely when there exists a bipartite input state that, when acted upon by \( \higherfont{M} \) on one subsystem, is sent to an output operator that is not positive semidefinite and hence not a valid state.
Such an \( \higherfont{M} \) is therefore unsuitable for representing a physical process.
Indeed, imagine the input system of \( \higherfont{M} \) is entangled with another system \Sys{C}, unbeknownst to the experimenter.
We still want the overall output to be a valid state.
The insistence on considering the local action of a transformation is particularly important for a good theory of quantum information, given the notorious difficulty of isolating a quantum system from its environment, even under ideal laboratory conditions.

To recall the precise definition, a map \( {\higherfont{M} \in \LinearOps\big(\LinearOps(\HilSpace{\Sys{A}}), \LinearOps(\HilSpace{\Sys{B}})\big)} \) is \DefnFont{completely positive} if for all finite-dimensional Hilbert spaces \( \HilSpace{\Sys{C}} \) and all \( \rho \in \PositiveOps{\HilSpace{\Sys{A}} \otimes \HilSpace{\Sys{C}}} \), we have
\begin{equation}
  (\higherfont{M} \otimes \mathcal{I}_{\Sys{C} \to \Sys{C}})(\rho) \in \PositiveOps{\HilSpace{\Sys{B}} \otimes \HilSpace{\Sys{C}}} .
\end{equation}
Note that we write \( \PositiveOps{\HilSpace{\Sys{B}} \otimes \HilSpace{\Sys{C}}} \) for the subset of positive semidefinite operators within \( {\LinearOps(\HilSpace{\Sys{B}} \otimes \HilSpace{\Sys{C}})} \),
and that we write \( \mathcal{I}_{\Sys{C} \to \Sys{C}} \) for the identity map of type \( \Sys{C} \tyto \Sys{C} \).

In ~\cite{chiribellaTransformingQuantumOperations2008,gourComparisonQuantumChannels2019,burnistonNecessarySufficientConditions2020}, the authors extended the same logic to the analysis of superchannels.
They argued that a superchannel must be completely CP-preserving (or, ``CCPP'') because, otherwise, such a map would send some valid bipartite input channel to a non-CP (thus invalid) output when applied to only one part.
The same logic can be extended yet another step upward,
resulting in the notion of completely CCPP-preserving (or, ``CCCPPP'') maps,
and yet another step, resulting in completely CCCPPP-preserving maps, and so on indefinitely.
We introduce the \KMaps family to define all these properties with a single inductive construction, and in a way that avoids the problem of exploding terminology exemplified by the terms ``completely CP-preserving,'' ``completely \mbox{CCPP-preserving},'' etc.

As we now see, the definition of the \KMaps family is stated with a straightforward inductive formula.
For elementary types \( x \), \( \KMaps(x) \) is the subset of positive semidefinite operators \( \PositiveOps{\HilSpace{x}} \subset \LinearOps(\HilSpace{x}) \) (\textit{caveat}, note the remark about the trivial type below).
For a non-elementary type \( x = a \tyto b \),
\( \KMaps(a \tyto b) \) is the subset of \DefnFont{completely} \KMaps-\DefnFont{preserving maps}.
Note that for \( z \in \ty \), we write \( \id{z\tyto z} \) for the identity map on \( \LinearMaps(z) \), and that \( \id{z \tyto z} \) is of type \( z \tyto z \).

\begin{definition} \label{def:completelypositivemaps}
  Let \( x \in \ty \) and \( \higherfont{M} : x \). Let \( \KMaps(x) \subset \LinearMaps(x) \) be defined by the following logical equivalence:
  \begin{equation} \label{eq:positivepreservingmaps1}
    \higherfont{M} \in \KMaps(x) \ \Longleftrightarrow \
    \begin{cases*}
       \higherfont{M} \in \PositiveOps{\HilSpace{x}}  & if \( x \in \et \) \\
      \higherfont{M} \text{ is completely } \KMaps\text{-preserving} & if \( x = a \tyto b \)
    \end{cases*}
    ,
  \end{equation}
  where a map \( \higherfont{M} : a \tyto b \) is called \DefnFont{completely} \KMaps-\DefnFont{preserving} if
  \begin{equation}
    (\higherfont{M} \parmap \mathcal{I}_{z \to z})(\rho)\in \KMaps(b\partype z)
  \end{equation}
  for all \( z\in\ty \) such that \( \tyord{z \tyto z} = \tyord{x} \) and for all \( \rho \in \KMaps(a \partype z) \).
\end{definition}
Note that in the case of the trivial type, we write \( \PositiveOps{ \HilSpace{\Sys{I}} } \) for the subset \( \mathbb{R}_{\geq0} \subset \mathbb{C} \) of positive real numbers.

Let us discuss the validity of Definition~\ref{def:completelypositivemaps} as an inductive construction.
In the case of \( x = a \tyto b \),
the definition of \( \KMaps(x) \) depends on the definition of \( \KMaps(a \partype z) \) and \( \KMaps(b \partype z) \) for every type \( z \) satisfying a certain condition.
Crucially, the assumptions guarantee that \( a \partype z \) and \( b \partype z \) are necessarily of order strictly less than the order of \( x \), and
consequently, the construction is valid by induction on the order of \( x \).
To conclude this discussion, we mention how to prove that \( \tyord{a \partype z} \) and \( \tyord{b \partype z} \) are strictly less than \( \tyord{x} \).
By Lemma~\ref{lem:parallelproductorder}, \( {\tyord{y \partype y'} = \max\{\tyord{y}, \tyord{y'}\}} \).
Therefore, given \( \tyord{z \tyto z} = \tyord{x} \) we can deduce that \( \tyord{x \partype (z \tyto z)} = \tyord{x} \).
It follows immediately that both \( \tyin{x \partype (z \tyto z)} = a \partype z \) and \( \tyout{x \partype (z \tyto z)} = b \partype z \) are of order strictly less than \( x \), because by definition we have
\begin{equation}
  \tyord{x \partype (z \tyto z)} \ = \ 1 + \max\{ \tyord{\tyin{x \partype (z \tyto z)}}, \tyord{\tyout{x \partype (z \tyto z)}} \} .
\end{equation}

In \cite{zanoniMapsHigherOrderMaps2025,steakleyTypebasedFrameworkHigherorder} and in forthcoming work \cite{steakleyOperationalHigherorderQuantum},
we demonstrate that completely \KMaps-preserving maps are closed under both sequential composition and parallel product.

\subsection{ \textsf{K} as a family of cones and the Choi isomorphism} \label{sec:propertiesofK}


Recall that a convex (linear) cone in a real vector space \( V \) is a subset \( \arbcone \subset V \) which is closed under non-negative linear combinations.
For instance, for any finite-dimensional Hilbert space \( \HilSpace{\Sys{A}} \), the Hermitian operators \( \HermitianOps{\HilSpace{\Sys{A}}} \) form a real vector space, and the set of positive semidefinite operators \( \PositiveOps{\HilSpace{\Sys{A}}} \subset \HermitianOps{\HilSpace{\Sys{A}}} \) is a convex cone.

To show that \( \KMaps(x) \) is a convex cone, we need to situate it inside a suitable real vector space.
For this, we appeal to a family of real vector spaces \( \HMaps(x) \subset \LinearMaps(x) \), for each \( x \in \ty \), which is defined in a way quite similar to the \KMaps family.
The difference is that for a non-elementary type \( x = a \tyto b \),
\( \HMaps(x) \subset \LinearMaps(x) \) is the subset of \HMaps-preserving maps (as opposed to \emph{completely} \HMaps-preserving).

\begin{definition} \label{def:Hermitianpreservingmaps}
  Let \( x \in \ty \) and \( \higherfont{M} : x \). Let \( \HMaps(x) \subset \LinearMaps(x) \) be defined by the following logical equivalence:
  \begin{equation} \label{eq:Hermitianpreservingmaps}
    \higherfont{M} \in \HMaps(x) \ \Longleftrightarrow \
    \begin{cases*}
      \higherfont{M} \in \HermitianOps{\HilSpace{x}}  & if \( x \in \et \) \\
      \higherfont{M} \text{ is } \HMaps\text{-preserving} & if \( x = a \tyto b \)
    \end{cases*}
    ,
  \end{equation}
  where a map \( \higherfont{M} : a \tyto b \) is called \HMaps-\DefnFont{preserving} if \( \higherfont{M}(\rho) \in \HMaps(b) \) for all \( \rho \in \HMaps(a) \).
\end{definition}
We omit the modifier ``completely'' not because we are deemphasizing local application of transformations, but because in the case of the \HMaps family, we are able to show that there is a logical equivalence between the notions of \HMaps-preserving map and completely \HMaps-preserving map \cite{steakleyOperationalHigherorderQuantum,steakleyTypebasedFrameworkHigherorder,zanoniMapsHigherOrderMaps2025}.

Note that in the case of the trivial type, we write \(  \HermitianOps{\HilSpace{\Sys{I}}}  \) for the subset \( \mathbb{R} \subset \mathbb{C} \) of real numbers.

It is relatively straightforward to show that $\HMaps(x)$ is a real vector space for any type $x$ (cf.\ Proposition~\ref{prop:real vector spaces}).
To derive further facts about $\HMaps(x)$ and about its relationship with $\KMaps(x)$, we appeal to a version of the Choi isomorphism based on our parallel product operation.
The main idea of the standard Choi isomorphism (for a typical definition, see \cite{bisioTheoreticalFrameworkHigherorder2019}) is to unitarily convert a map of type \( \Sys{A} \tyto \Sys{B} \) into a matrix of type \( \Sys{BA} \).
Our version extends this to maps of all orders.

Fix an orthonormal basis \( \{ u_j \} \) of \( \LinearMaps(a) \).
Given a map \( \higherfont{M} : a \tyto b \), define
\begin{equation}
  \Choi_{a \tyto b}(\higherfont{M}) := \sum_{j} \higherfont{M}(u_j) \parmap u_j .
\end{equation}
\( \Choi_{a \tyto b} (\higherfont{M}) \) is then a linear map of type \( b \partype a \), so $\Choi_{a \tyto b}$ decreases the order of $\higherfont{M}$ by one. The assignment \( \higherfont{M} \mapsto \Choi_{a \tyto b}(\higherfont{M}) \) defines a linear (in fact, unitary) map of type \( (a \tyto b) \tyto b \partype a \).

In \cite{steakleyOperationalHigherorderQuantum,steakleyTypebasedFrameworkHigherorder},
we show that chaining together multiple instances of the Choi isomorphism results in an 
isomorphism of convex cones between each \( \KMaps(x) \) and a cone of positive semidefinite operators \( \PositiveOps{\MyH} \), for a suitably chosen finite-dimensional Hilbert space \MyH.
Since these cones are isomorphic, properties of \( \PositiveOps{\MyH} \) transfer to \( \KMaps(x) \). For instance, the property that \( \PositiveOps{\MyH} \) spans \(  \HermitianOps{\MyH}  \) translates to the fact that \( \KMaps(x) \) spans \( \HMaps(x) \), which is to say that for every \( \higherfont{M} \in \HMaps(x) \), there exist maps \( \higherfont{M}_+,\higherfont{M}_- \in \KMaps(x) \) such that \( \higherfont{M} = \higherfont{M}_+ - \higherfont{M}_- \).

The property of self-duality, which plays a useful role in conic linear programs for optimization, also transfers to \( \KMaps(x) \) from \( \PositiveOps{\MyH} \).
The proofs of these facts are rather involved, and so we refer the reader to our more detailed results \cite{steakleyOperationalHigherorderQuantum,zanoniMapsHigherOrderMaps2025,steakleyTypebasedFrameworkHigherorder}.

%




As a further consequence of the isomorphism of cones \( \KMaps(x) \cong \PositiveOps{\MyH} \) provided by the Choi isomorphism,
we can fairly easily check if a map is completely \KMaps-preserving, even though Definition~\ref{def:completelypositivemaps} seems to suggest that it is a daunting task. In this respect, it is actually enough to check if the chain of successful applications of Choi isomorphisms produces a positive semidefinite operator.

Therefore,  the \KMaps family, a generalization of the positive semidefinite cone to higher orders, preserves the geometric properties of such a cone to all orders.
This family is also a crucial step towards characterizing the physically meaningful maps at any order.
The other aspects of such a characterization, e.g.\ the higher-order generalization of trace-preservation, will be examined in future work.

\section{Conclusions} \label{sec:conclusion}

\nocite{skrzypczykSemidefiniteProgrammingQuantum2023}
\nocite{}

We have introduced a framework for higher-order quantum theory in the finite-dimensional case.
The basic components of the framework are a simple type system for describing different kinds of higher-order maps and a Hilbert space of linear maps for each type.
These spaces include all forms of higher-order quantum protocols.
In addition to the usual sequential composition of linear maps, we have equipped the framework with a parallel compositional structure in the form of the parallel product operation.
From this we gain an operation that fulfills in full generality the role that the tensor product plays for pairs of states and pairs of channels, namely providing an operationally meaningful notion of parallel combination of maps.

The inductive constructions at the heart of our framework, which are made possible by its type system, provide us with economical notation and terminology
in a context where economy is much needed, due to the infinitely many different types of object that higher-order quantum theory inevitably involves.
For instance, we can write \( \LinearMaps( (\Sys{A} \tyto \Sys{B} ) \tyto (\Sys{C} \tyto \Sys{D}) ) \) instead of a cumbersome expression such as \( \LinearOps\Big( \LinearOps\big( \LinearOps(\HilSpace{\Sys{A}}) , \hspace{0.1em} \LinearOps(\HilSpace{\Sys{B}}) \big) , \hspace{0.1em} \LinearOps\big( \LinearOps(\HilSpace{\Sys{C}}) , \hspace{0.1em} \LinearOps(\HilSpace{\Sys{D}}) \big) \Big) \),
and we can say ``completely \KMaps-preserving maps'' instead of ``completely CP-preserving maps,'' ``completely CCPP-preserving maps,'' etc.
The type system also facilitates the use of powerful inductive constructions that are based on simple principles,
such as when we defined the parallel product operations for types and maps.

The definitive point of departure of our framework in comparison to previous work \cite{bisioTheoreticalFrameworkHigherorder2019,kissingerCategoricalSemanticsCausal2019} is the methodological decision to avoid using the Choi representation to define the framework's basic objects.
However, because the Choi representation derives from certain linear maps between Hilbert spaces, it can be defined within our framework (for details, see \cite{steakleyOperationalHigherorderQuantum,steakleyTypebasedFrameworkHigherorder,zanoniMapsHigherOrderMaps2025}).
By introducing the Choi representation itself from our framework, rather than incorporating it into the formalism at the very foundation,
we can avoid notational ambiguities that can arise when using Choi matrices in calculations, as noted in \cite{zanoniChoidefinedResourceTheories2025}.
And because the Choi representation is used only where it is strictly pertinent,
it promotes the ``decoupling'' of the study of higher-order quantum theory from the Choi formalism, paving the way to an extension to the infinite-dimensional case.


After setting up the framework's methodological foundation, we must take up the characterization of the physically meaningful objects as a subset of \( \LinearMaps(x) \) for each \( x \in \ty \).
In this work, we addressed the aspect of complete positivity.
Thanks to the parallel product operation, the \KMaps family can be constructed using a straightforward inductive formulation of the notion of ``completely \mbox{\KMaps-preserving} map,'' which generalizes the notion of complete positivity.
Despite the generalization,
it remains the case that \( \KMaps(x) \) is a convex cone, and is furthermore isomorphic to a cone of positive semidefinite operators, thus sharing its many nice properties.
Consequently, many of the same convex optimization techniques that can be used to solve problems concerning states and channels \cite{girardConvexOptimizationProblems2014,skrzypczykSemidefiniteProgrammingQuantum2023} are applicable to these higher-order maps as well.
The construction of the \KMaps family also represents a first test of the suitability of the parallel product operation to occupy a basic role in the mathematics of higher-order quantum theory.

It remains for future work to complete the full characterization of the physically meaningful objects, which should form a subset of \( \KMaps(x) \) for each type \( x \).
In keeping with the precedent set by previous works, \cite{bisioTheoreticalFrameworkHigherorder2019,kissingerCategoricalSemanticsCausal2019}, the next step would be to develop an appropriate definition of \emph{deterministic} map for all types.
The full higher-order definition would generalize states, channels, and superchannels, which are the deterministic maps in the familiar low-order cases.
One approach, similar in spirit to \cite{bisioTheoreticalFrameworkHigherorder2019} and \cite{kissingerCategoricalSemanticsCausal2019}, would be to define a general deterministic transformation inductively, by the requirement to send deterministic input maps to deterministic output maps, potentially in a complete sense with respect to our parallel product.

Once we have a suitable characterization of the physically meaningful maps, 
we intend to use our framework to study quantum processes that exhibit indefinite causal order \cite{chiribellaQuantumComputationsDefinite2013,oreshkovQuantumCorrelationsNo2012,oreshkovCausalCausallySeparable2016,baumelerSpaceLogicallyConsistent2016,wechsQuantumCircuitsClassical2021}.
Such processes should be included by a permissive characterization of physically meaningful maps.
We aim to investigate whether processes with indefinite causal order are associated with any special signature within our framework.
Another direction we intend to pursue is the extension of quantum information techniques to higher-order transformations, such as the generalization of the notion of channel entropy \cite{gourComparisonQuantumChannels2019,gourEntropyQuantumChannel2021,gourInevitableNegativityAdditivity2025a,yuanHypothesisTestingEntropies2019,chuEntropyFunctionQuantum2022}.
This problem could be approached one order at a time, first passing from channels to superchannels, but if an adequate inductive characterization can be found then it could be solved at once for all orders.
Another potential area of application, and one of the main motivations for this work, is quantum resource theories \cite{chitambarQuantumResourceTheories2019,gourQuantumResourceTheories2025,Das2024,Sohail2026,Badhani2026,Das2026,Badhani2026a}.
In extant work, the resources studied are either static (states) or dynamical (channels), but it has been shown that higher-order processes like the SWITCH can also convey a quantum advantage in information processing \cite{colnaghiQuantumComputationProgrammable2012, chiribellaPerfectDiscriminationNosignalling2012,chiribellaQuantumShannonTheory2019,kristjanssonResourceTheoriesCommunication2020,kristjanssonSecondquantisedShannonTheory2022,zhaoQuantumMetrologyIndefinite2020}.
To expand the scope to include all higher-order processes would be to open a broad new domain for the general theory of quantum resources, in the hope of uncovering new sources of quantum advantage potentially present at higher orders.

\paragraph{Data availability statement}
This article has no associated data.	

\paragraph{Conflict of interest statement}
The authors have no conflicts of interest.

\paragraph{Funding statement}
E.\ Z.\ acknowledges support from the Eric Milner’s Graduate Scholarship, the Alberta Graduate Excellence Scholarship (AGES), and the Eyes High International Doctoral Scholarship. C.\ M.\ S.\ acknowledges the support of the Natural Sciences and Engineering Research Council of Canada (NSERC) through the Discovery Grant “The power of quantum resources” RGPIN-2022-03025 and the Discovery Launch Supplement DGECR-2022-00119.

\appendix

\section{Additional proofs} \label{app}

In this appendix, for the interested reader we reproduce some proofs of results mentioned in the main body of this article (see also~\cite{steakleyOperationalHigherorderQuantum,steakleyTypebasedFrameworkHigherorder,zanoniMapsHigherOrderMaps2025}). We start from the proof of Lemma~\ref{lem:uniquereadability}, which describes the form of a generic type.

\begin{proof}[Proof of Lemma~\ref{lem:uniquereadability}]
  Suppose that \( x \notin \et \) and that there do not exist \( a,b \in \ty \) such that \( x = a \tyto b \).
  Then \( x \) could be removed from \ty
  and the resulting set, \( \ty \setminus \{x\} \), would still satisfy \eqref{eq:typesrule1} and \eqref{eq:typesrule2},
  contradicting the definition of \ty as the \emph{smallest} such set (with respect to subset inclusion).
  In conclusion, either \( x \in \et \) or there exist \( a,b \in \ty \) such that \( x = a \tyto b \).
  Both cannot be true at the same time, because elements of \et do not contain the arrow symbol.
  
  Now it suffices to show that if \( x = a \tyto b \), then \( a \) and \( b \) are unique as such.
  Suppose \( a',b'\in\ty \) and \( a \tyto b = a' \tyto b' \).
  Then as sequences of symbols, \( a \tyto b \) and \( a' \tyto b' \) must have the same number of symbols, and their symbols must match at every position.
  Recall the rules we gave governing how to add brackets when forming an expression such as \( a \tyto b \), and notice that they guarantee that any such type contains exactly one arrow symbol that does not lie between any pair of brackets.
  Call this the outer arrow.
  We know that the outer arrow of \( a \tyto b \) occurs at the same position as the outer arrow of \( a' \tyto b' \).
  Furthermore, if we take the subexpression of all symbols lying to the left of the outer arrow, this subexpression must be the same for both \( a \tyto b \) and \( a' \tyto b' \).
  If this subexpression is not wrapped in brackets, then for \( a \tyto b \), the subexpression is exactly \( a \), while for \( a' \tyto b' \) it is \( a' \), and so we deduce that \( a = a' \).
  Otherwise, if the subexpression is wrapped in brackets, we obtain the equation of expressions \( (a) = (a') \), which also implies that \( a = a' \). Notice that we can never have the situation in which only one between $a$ and $a'$ is wrapped in brackets, otherwise we could not have the equality \( a \tyto b = a' \tyto b' \).
  The same argument applies to the subexpression of all symbols to the right of the outer arrow, in which case we deduce \( b = b' \).
\end{proof}

Now we move the proof of Lemma~\ref{lem:parallelproductorder}, which describes the order of the parallel product of two types.

\begin{proof}[Proof of Lemma~\ref{lem:parallelproductorder}]
  By induction on \( \InductionIndex = \max\{\tyord{x},\tyord{y}\} \).
  If \( \InductionIndex = 0 \), then \( x,y \) are both elementary and so is \( x \partype y \).
  The proposition follows because all elementary types have order zero.

  Let \( \InductionIndex > 0 \), and assume as induction hypothesis that Eq.~\eqref{eq:paralleltypeorder} holds for all pairs of types \( x',y' \) of lower order (i.e.\ less than \(\InductionIndex \)).
  The assumption \( \InductionIndex > 0 \) implies that either \( x \) or \( y \) is non-elementary.
  Write \( x = a \tyto b \) if \( x \) is non-elementary, or write \( y = c \tyto d \) if \( y \) is.
  So by Definition~\ref{def:typeorder} and Definition~\ref{def:paralleltype}, we have
  \begin{equation} \label{eq:paralleltypeorder1}
    \tyord{x \partype y} = \begin{cases*}
      1 + \max\{\tyord{c}, \tyord{x \partype d}\} & if \( \tyord{x} < \tyord{y} \) \\
      1 + \max\{\tyord{a \partype c}, \tyord{b \partype d}\} & if \( \tyord{x} = \tyord{y} \) \\
      1 + \max\{\tyord{a}, \tyord{b \partype y}\} & if \( \tyord{x} > \tyord{y} \)
    \end{cases*}
  \end{equation}
  For each expression of the form \( \tyord{x' \partype y'} \) on the right-hand side of Eq.~\eqref{eq:paralleltypeorder1}, it is guaranteed by the surrounding assumptions that the types \( x' \) and \( y' \) are of order less than \( \InductionIndex \).
  Consequently, we can apply the induction hypothesis to these expressions \( \tyord{x' \partype y'} \) to find that
  \begin{equation} \label{eq:paralleltypeorder2}
    \tyord{x \partype y} =
    \begin{cases*}
      1 + \max\{\tyord{c}, \tyord{x}, \tyord{d}\} & if \( \tyord{x} < \tyord{y} \) \\
      1 + \max\{\tyord{a}, \tyord{c}, \tyord{b}, \tyord{d}\} & if \( \tyord{x} = \tyord{y} \) \\
      1 + \max\{\tyord{a}, \tyord{b}, \tyord{y}\} & if \( \tyord{x} > \tyord{y} \)
    \end{cases*}
  \end{equation}
  If \( \tyord{x} < \tyord{y} \), then 
  \begin{equation}
    1 + \max\{\tyord{c}, \tyord{x}, \tyord{d}\} = 1 + \max\{\tyord{c}, \tyord{d}\} = \tyord{y} = \max\{\tyord{x}, \tyord{y}\} .
  \end{equation}
  Similarly, \( 1 + \max\{\tyord{a}, \tyord{b}, \tyord{y}\} = \max\{\tyord{x}, \tyord{y}\} \) if \( \tyord{y} < \tyord{x} \).
  Otherwise, \( \tyord{x} = \tyord{y} \) and we have
  \begin{equation}
    1 + \max\{\tyord{a},\tyord{c},\tyord{b},\tyord{d}\} \ = \ 1 + \max\{\tyord{x},\tyord{y}\} - 1 \ = \ \max\{\tyord{x},\tyord{y}\}.
  \end{equation}
\end{proof}

Finally, we  prove that the $\mathsf{H}$ family introduced in Section~\ref{sec:propertiesofK} is made of real Hilbert spaces.

\begin{proposition}\label{prop:real vector spaces}
  For all \( x \in \ty \), \( \mathsf{H}(x) \) is a real Hilbert space, with inner product given by the Hilbert-Schmidt inner product.
\end{proposition}
\begin{proof}
  By induction.
  When \( x = \mathrm{I} \), the proposition is that \( \mathbb{R} \) is a real vector space and the Hilbert-Schmidt inner product in this case is equivalent to the product of real numbers. If $x\neq \mathrm{I}$ is a non-trivial elementary type, $\HMaps(x)=\HermitianOps{\HilSpace{x}} $, which is a real vector space naturally endowed with the Hilbert-Schmidt inner product. 

  Let \( x = a \tyto b \) and suppose by induction that the proposition holds for \( x = a \) and \( x = b \).
  To show that \( \HMaps(x) \) is a real vector space, it suffices to show that \( \mathcal{M} + \mathcal{N} \) and \( r \mathcal{M} \) are both \( \HMaps \)-preserving for any \( \mathcal{M},\mathcal{N} \in \HMaps(x) \) and any \( r \in \mathbb{R} \).
  Given any \( \rho \in \HMaps(a) \), we calculate:
  \begin{align}
    (\mathcal{M} + \mathcal{N})(\rho) & = \mathcal{M}(\rho) +\mathcal{N}(\rho) \\
    (r \mathcal{M})(\rho) & = r  \mathcal{M}(\rho)
  \end{align}
  Since \( \mathcal{M}\) and $\mathcal{N}$ are \( \HMaps \)-preserving, and \( \HMaps(b) \) is a real vector space by induction hypothesis, the right-hand sides of both the above equations are elements of \( \HMaps(b) \).

  Now it is left to demonstrate that the Hilbert-Schmidt inner product is symmetric on \( \HMaps(a \tyto b) \).
  The proof relies on the fact that there is an orthonormal basis of \( \LinearMaps(x) \) that lies entirely within \( \HMaps(x) \)
  for every \( x \in \ty \) \cite{zanoniMapsHigherOrderMaps2025,steakleyTypebasedFrameworkHigherorder}. 
We assume as induction hypothesis that
  the Hilbert-Schmidt inner product is symmetric on \( \HMaps(b) \),
  and that we have a collection of maps
 \( \{ u_j \} \) that forms an orthonormal basis of both \( \LinearMaps(a) \) and \( \HMaps(a) \).
  The first step is to expand the inner product \( \LeftIP \mathcal{N} , \mathcal{M} \RightIP_x \) according to the formula that computes traces from an orthonormal basis:
  \begin{equation}
    \LeftIP \mathcal{N} , \mathcal{M} \RightIP_x =
    \Trace{\Adj{\mathcal{N}} \circ \mathcal{M}} =
    \sum_{j} \big\LeftIP u_j , \hspace{.1em} \Adj{\mathcal{N}} \circ \mathcal{M} (u_j) \big\RightIP_a
  \end{equation}
  By the defining properties of the adjoint, \( \LeftIP u_j , \Adj{\mathcal{N}} \circ \mathcal{M}(u_j) \RightIP_a \) is equal to \( \LeftIP \mathcal{N} (u_j) , \mathcal{M}(u_j) \RightIP_b \).
  Thanks to the induction hypothesis, we can apply symmetry of the Hilbert-Schmidt inner product on maps of type \( b \).
  Then, by applying the previous steps in reverse order, we have
  \begin{align}
    \LeftIP \mathcal{N} , \mathcal{M} \RightIP_x & =
    \sum_{j} \big\LeftIP \mathcal{M} (u_j) , \mathcal{N} (u_j) \big\RightIP_b \\
    & = \Trace{\Adj{\mathcal{M}} \circ \mathcal{N}} \\
    & = \LeftIP \mathcal{M} , \mathcal{N} \RightIP_x.
  \end{align}
\end{proof}

\bibliography{higherorderlibrary}

\end{document}